\documentclass[a4paper,11pt]{article}
\usepackage{jheppub} 
\usepackage{lineno}
\nolinenumbers
\usepackage{amsmath}
\usepackage{mathrsfs}
\usepackage{amsfonts}

\title{\boldmath Semi-Universality of $O(N)$ model on $S^1\times S^2$}

\author{Vinayak Mishra}
\affiliation{Department of Theoretical Physics,\\
Tata Institute of Fundamental Research, Mumbai-400005, India}

\emailAdd{vinayak.mishra@tifr.res.in}

\abstract{We study the thermal partition function of the critical large-$N$ $O(N)$ model on rotating $S^1\times S^2$. Rotation makes the saddle latitude dependent, with its leading profile governed by the local temperature. Using a Weyl-covariant hydrostatic expansion, we determine the partition function through four-derivative order in high-temperature and semi-universal limits. In the near-light-speed limit, the geometry around the equator reduces to a pp-wave and captures the expected semi-universal behavior of the partition function. We compute the corresponding theory-dependent residue directly on this limiting geometry, both at high and low temperatures. The high-temperature result agrees with the near-light-speed limit of the sphere calculation, while the low-temperature expansion reveals interaction-dependent corrections arising from the response of the saddle to the thermal source.
}

\begin{document}
\maketitle
\flushbottom

\section{Introduction}\label{s:int}

The study of conformal field theories (CFTs) has been of longstanding interest, with local operators playing a central role in characterizing any CFT. The thermal partition function of a CFT on $S^1\times S^{d-1}$ encodes the spectrum of local operators and is therefore a natural object of interest. Although the partition function is in general theory dependent, it exhibits certain universal structures in appropriate limits.

We are concerned with the thermal partition function on $S^1\times S^2$, with angular velocity $\omega$ on $S^2$ and inverse temperature $\beta$ being the periodicity of $S^1$. The canonical partition function is defined by
\begin{equation}\label{eq:cpf}
Z(\beta,\omega)=\operatorname{Tr}e^{-\beta(H-\omega J)}\quad |\omega|<1.
\end{equation}
Here $H$ is the Hamiltonian of the theory and $J$ is the angular momentum associated with rotation on $S^2$. For $\omega=0$, the partition function \eqref{eq:cpf} exhibits the universal high-temperature behavior $\log Z\propto T^2$. This follows from dimensional analysis together with the extensivity of the partition function; more generally, in $d$ spacetime dimensions one expects $\log Z\propto T^{d-1}$. This behavior was generalized to non-zero $\omega$ in \cite{Bhattacharyya:2007vs}, where the rotating partition function was also shown to exhibit a universal structure. More recently, it was conjectured in \cite{Anand:2025mfh} that, at finite temperature, the CFT partition function develops a universal pole structure in the limit $\omega\to1$. A closely related feature also appears in the Grey-Galaxy partition functions discussed in \cite{Bajaj:2024utv,Kim:2023sig}. The universal behavior conjectured in \cite{Anand:2025mfh} takes the form
\begin{equation}\label{eq:smpf}
\log Z(\beta,\omega)=\frac{1}{1-\omega^2}h(\beta)\big(1+O(1-\omega)\big),
\qquad
\omega\to1,\quad \beta\ \text{fixed}.
\end{equation}
Here the pole structure is universal, whereas the residue $h(\beta)$ is theory dependent, leading to the notion of \emph{semi-universality}. This phenomenon was subsequently given a geometric interpretation in \cite{Komargodski:2026ain}, where the pole in \eqref{eq:smpf} was understood as arising from an emergent volume in an appropriate limit of $S^1\times S^2$. The behavior in \eqref{eq:smpf} was also verified for a variety of free theories in \cite{Mukherjee:2026dfu}.

To compute the partition function, we make essential use of the equilibrium partition-function formalism developed in \cite{Banerjee:2012iz,Jensen:2012jh}, which in more recent language may be viewed as a thermal Effective theory \cite{Benjamin:2023ThermalEFT} (also see:~\cite{Allameh:2024qqp}) . The large-$N$ critical $O(N)$ model, or equivalently the large-$N$ Wilson--Fisher fixed point, is an exactly solvable CFT in $d=3$ admitting a systematic large-$N$ expansion \cite{PhysRevD.7.2911}; see \cite{Moshe:2003xn} for a review. The theory was studied at zero temperature in flat space using a Hubbard--Stratonovich (HS) auxiliary field in \cite{PhysRevD.10.2491,PhysRevD.13.2212}. Its finite-temperature formulation was developed along similar lines in \cite{ChubukovSachdevYe:1994}; see in particular part~III on the quantum $O(N)$ non-linear sigma model. The free energy and mass gap in flat space were computed in \cite{Sachdev:1993pr}. The thermodynamics of the $O(N)$ model was further studied in \cite{Romatschke:2019ybu}, providing useful consistency checks across the full range of couplings.

From a more conventional CFT perspective, \cite{Petkou:1998fc} studied the theory using the operator-product expansion, while \cite{Iliesiu:2018fao} analyzed its thermal correlators through the finite-temperature bootstrap program. The thermal correlation functions and free energy on $S^1\times S^2$ were subsequently studied in \cite{David:2024pir,David:2025tqn}. More recently, \cite{Mauro:2026zus} computed the Effective action of the theory on a rotating $S^2$ and obtained small-$\omega$ corrections to the free energy previously computed in \cite{David:2024pir}. The rotating problem on $S^1\times S^2$ presents a distinctive complication, that in the non-rotating case the HS saddle, can be taken to be constant, whereas rotation makes the local thermal environment latitude dependent. We emphasize this point and show that, once rotation is included, the HS field is no longer constant but develops a non-trivial dependence on the latitude. This distinction is particularly relevant to the recent work of \cite{David:2026yis}, which studies a related vector model whose Hubbard--Stratonovich field is restricted to its spatially constant mode. As explained there, this restriction defines a different model from the local critical $O(N)$ theory at nonzero angular velocity; the nonuniformity of the saddle in the local theory is also demonstrated in their appendix D. In the present work, developed independently of that study, we retain the latitude-dependent auxiliary field and solve the local gap equation perturbatively in the regimes considered. The two calculations therefore address distinct saddle-point problems, and their thermodynamic results need not coincide at nonzero angular velocity.

\subsection{Results and Summary}

We compute the leading large-\(N\) thermal partition function of the critical
\(O(N)\) model on the rotating background \(S^1\times S^2\), setting the
radius of the sphere to unity. Rotation is introduced through the thermal
identification
\begin{equation} (\tau,\theta,\varphi) \sim (\tau+\beta,\theta,\varphi-i\beta\omega), \qquad |\omega|<1 .
\end{equation}

\textbf{Latitude-dependent Saddle}

In contrast to the non-rotating theory, the Hubbard--Stratonovich saddle is not
spatially constant. We find the leading high-temperature saddle
\begin{equation}
    \bar\sigma(\theta)=\frac{4\log^2\varphi_{ g}}{\beta^2(1-\omega^2\sin^2\theta)} +O(\beta^0),
\end{equation}
here $\varphi_g$ is the golden ratio $\left(\varphi_g=\frac{1+\sqrt{5}}{2}\right)$.
Thus, the dynamically generated mass squared is proportional to the square of
the local Tolman temperature \cite{Tolman:1930gr}. In particular, the saddle is increasingly
enhanced near the equator as \(\omega\to1\).

\textbf{High-Temperature partition function on $S^1\times S^2$}

Using a Weyl-covariant derivative expansion of the hydrostatic Effective
action, together with a heat-kernel determination of its Wilson coefficients, 
we obtain an fixed $\omega$ but high-temperature $(\beta\to0)$ asymptotic expression for $\log Z$
\begin{equation}\label{eq:ispf}
\begin{aligned}
    \log Z(\beta,\omega)=
    \frac{2\pi N}{1-\omega^2}\Bigg[&\frac{4\zeta(3)}{5\pi\beta^2}
        -\frac{\sqrt{5}\log\varphi_{ g}}{18\pi}\,\omega^2\\[2mm]
        &+\frac{\beta^2}{10800\pi}\Bigg\{
            \frac{15\sqrt{5}}{2\log\varphi_{ g}}-20\left(1+\frac{\sqrt{5}}{2\log\varphi_{ g}}
            \right)\omega^2\\
        &\hspace{32mm}
            +\left(46+\frac{13\sqrt{5}}{2\log\varphi_{ g}}+60\sqrt{5}\log\varphi_{ g}\right)\omega^4\Bigg\}+O(\beta^4)\Bigg].
\end{aligned}
\end{equation}
For \(\omega=0\), equation~\eqref{eq:ispf}
reproduces the known high-temperature expansion on \(S^1_\beta\times S^2\).
Its expansion to quadratic order in \(\omega\) also agrees with the previously
obtained small-angular-velocity result.

\textbf{Semi-universal limit}

We next study the near-light-speed limit by introducing
\begin{equation}
    \varepsilon\equiv1-\omega^2,\;\;
    \theta=\frac{\pi}{2}+\sqrt{\varepsilon}\,y,\;\;\phi=\varepsilon u,\;\;\varepsilon\rightarrow0 .
\end{equation}
After extracting an overall Weyl factor, the geometry near the hot equator
reduces to the pp-wave
\begin{equation}
    ds_{\rm{pp}}^2=(1+y^2)d\tau^2+dy^2-2i\,d\tau\,du .
\end{equation}
The partition function consequently develops a simple pole,
\begin{equation}
    \log Z(\beta,\omega)=\frac{h(\beta)}{1-\omega^2}\big(1+O(1-\omega)\big),\quad
    \omega\rightarrow1, \;\;\text{fixed}\;\beta
\end{equation}
where the theory-dependent residue \(h(\beta)\) is determined by the
partition-function density of the critical \(O(N)\) model on the pp-wave.

\textbf{Semi-universal residue at High-Temperature}

A direct worldline calculation gives the high-temperature asymptotics of residue
\begin{equation}
\begin{aligned}
    h(\beta)=2\pi N\Bigg[
        &\frac{4\zeta(3)}{5\pi\beta^2}-\frac{\sqrt{5}\log\varphi_{ g}}{18\pi}+c_2\beta^2\\
        &-0.0024762242577\,\beta^4
        +O(\beta^6)\Bigg],
\end{aligned}
\end{equation}
with
\begin{equation}
    c_2=\frac{13+\dfrac{2\sqrt{5}}{\log\varphi_{g}}+30\sqrt{5}\log\varphi_{ g}
    }{5400\pi}=0.003216943326\ldots .
\end{equation}
The analytic value of \(c_2\) agrees exactly with the
\(\omega\to1\) limit of the four-derivative result on the sphere. This provides
a nontrivial consistency check of the pp-wave reduction. The coefficient of $\beta^4$ was obtained numerically.

\textbf{Semi-universal residue at Low-Temperature}

Finally, expanding about the zero-temperature critical saddle, we obtain the
low-temperature asymptotics of residue
\begin{equation}
\begin{aligned}
    h(\beta)=2\pi N\Bigg[&\frac{e^{-\beta/2}}{\pi\beta}+\left(\frac{1}{4\pi\beta}-\frac{2}{\pi^3}\right)e^{-\beta}\\[1mm]
        &+\left(\frac{10}{9\pi\beta}+\frac{4\beta-36\log 2}{3\pi^3}+\frac{28\zeta(3)}{\pi^5}\right)e^{-3\beta/2}
        +O(e^{-2\beta})\Bigg].
\end{aligned}
\end{equation}
The leading scale \(e^{-\beta/2}\) is fixed by the lower edge of the pp-wave
single-particle spectrum. The subsequent terms contain both the free
dilute-gas contribution and the response of the interacting large-\(N\)
saddle to the thermal source.

\paragraph{Organization of paper}
In \S\ref{sec:cos1s2}, we formulate the critical \(O(N)\) model on
the rotating background \(S^1\times S^2\), construct its Weyl-covariant
hydrostatic Effective action, and determine the high-temperature saddle and
partition function through four-derivative order. In
\S\ref{sec:hote}, we take the near-light-speed limit and show that the
neighborhood of the hot equator reduces to a pp-wave geometry whose regulated
volume produces the \(1/(1-\omega^2)\) pole. In
\S\ref{sec:surmai}, we develop a worldline heat-kernel
representation of the pp-wave Effective action and derive the high- and
low-temperature expansions of its residue. In \S\ref{sec:diss}, we
discuss the implications of our results and possible extensions.
Appendix~\ref{ap:wifd} collects the Weyl-covariant fluid
dictionary, appendix~\ref{a:hksph} contains the calculation of
the local Wilson coefficients, appendix~\ref{app:pphk} presents
the worldline expansion, and appendix~\ref{ap:ld} provides
details of the low-temperature calculation.

\section{\texorpdfstring{Critical $O(N)$ model on $S^1\times S^2$}
{Critical O(N) model on S1 x S2}}
\label{sec:cos1s2}
We define the critical $O(N)$ Model with scalar field $\varphi^i$ where $i$ is $i=1....N$ as a flavor index, the Lagrangian Density of the model is given by
 \begin{equation}\label{eq:lag-dens}
\mathcal{L}=\frac12\partial_\mu\varphi^i\partial^\mu\varphi^i+\frac12\mu_0^2\varphi^i \varphi^i +\frac12\zeta_c R(g)\varphi^i\varphi^i+\frac{\lambda_0}{8N}(\varphi^i\varphi^i)^2
\end{equation}

Here, we have $\lambda_0$ as bare coupling, $\mu_0^2$ bare mass, $R(g)$ the Ricci Scalar of the Manifold and $\zeta_c=\frac{d-2}{4(d-1)}$ is the conformal coupling depending on dimensions of Manifold.

We linearize the quartic interaction term by  Hubbard-Stratonovich Transformation
\begin{equation}\label{eq:hs_transform}
        \begin{split}
            \mathcal{L}'&=\mathcal{L}-\frac{N}{2\lambda_0}\left(\sigma-\frac{\lambda_0}{2N}\varphi^i\varphi^i-\mu_0^2\right)^2\\
            &= \frac12(\partial_\mu\varphi^i)^2+\frac12 \zeta_cR(g)\varphi^i\varphi^i+\frac12\sigma\varphi^i\varphi^i-\frac{N}{2\lambda_0}\sigma^2+N\frac{\mu_0^2}{\lambda_0}\sigma
        \end{split}
\end{equation}

 We define the action of full theory on the manifold $S^1 \times S^2$ in the strong coupling limit of $\lambda_0\to\infty$\footnote{As usual in the large-\(N\) HS formulation, the auxiliary-field integral is defined by the contour inherited from the HS representation and evaluated by deformation to the appropriate saddle; see, e.g., \cite{Moshe:2003xn}. In the critical theory the \(\sigma^2/\lambda_0\) term is irrelevant and is dropped in the \(\lambda_0\to\infty\) limit \cite{Giombi:2019upv}.} and we ignore the constant $\mu_0^4$ term appearing in $\mathcal{L}'$, hence the euclidean action is 
\begin{equation}\label{eq:action}
        S_E[\Phi,\sigma]= \int d^3x\sqrt{g}\left[\frac12\varphi^i\left(-\nabla^2+\sigma+\frac18 R(g)\right)\varphi^i\right] + N\frac{\mu_0^2}{\lambda_0}\int d^3x \sqrt{g}\;\sigma,
\end{equation}
then the Effective action of the theory is given by 
\begin{equation}
\begin{split}
    e^{-S_{\text{eff}}(\sigma)}&= \int d\Phi \,e^{-S_E[\Phi,\sigma]}\\
    &= \exp\left\{-\frac{N}{2}\text{Tr}\ln\left(-\nabla^2+\sigma(x)+\frac{R}{8}\right)-N\frac{\mu_0^2}{\lambda_0}\int d^3x\sqrt{g}\,\sigma(x)\right\}.
\end{split}
\end{equation}
The bare mass $\mu_0$ represents a relevant deformation away from the critical theory; its value will be chosen (by fine tuning) in a manner that ensures that the field $\varphi^i$ are massless in flat space; this choice affects fine tuning to the critical point (setting the relevant deformation to zero). 

We define the function and central object of interest in this work as
\begin{equation}\label{eq:eff-action}
    \Gamma(\sigma)=\frac{1}{2}\text{Tr}\log\left(-\nabla^2+\sigma(x)+\frac{R}{8}\right)+\frac{\mu_0^2}{\lambda_0}\int d^3x\sqrt{g}\,\sigma(x),
\end{equation}
and the full partition function of the theory by
\begin{equation}\label{eq:sigmapi}
    Z = \int \mathcal{D}\sigma \;e^{-N\Gamma(\sigma(x))}.
\end{equation}
In the large $N$ limit, the path integral \eqref{eq:sigmapi} is evaluated in the saddle point approximation by solving the gap equation 
\begin{equation}\label{eq:gape}
    \frac{\delta\Gamma(\sigma)}{\delta\sigma(x)}\bigg|_{\bar\sigma(x)}=0.
\end{equation}
Let $\sigma=\bar\sigma (x)$ denote the dominant saddle point solution to this equation. It follows that 
\begin{equation}\label{eq:largeN}
    \ln Z = -N \Gamma\left(\bar\sigma(x) \right) + O(N^0),
\end{equation}

Now in order to define the critical theory, we fix the coefficients $\frac{\mu_0^2}{\lambda_0}$ on Quantum Critical Point $(T=0,\sigma=0)$, we are essentially following divergence removing scheme from \cite{Sachdev:1993pr}.

We have the manifold $S^1\times S^2$, with the metric
\begin{equation}\label{eq:metr}
    ds^2_{S^1\times S^2} = d\tau^2 +d\theta^2+\sin^2\theta d \phi^2,
\end{equation}
where $\tau\to\tau+\beta$ with $\beta$ being inverse temperature and periodicity of thermal circle $S^1$. The rotation on the $ S^2$ is introduced by identification  $(\tau,\theta,\phi)\to(\tau+\beta,\theta,\phi-i\omega\beta)$. This identification on the $\phi$ coordinate, consequently explicitly breaks the symmetry of $S^1\times S^2$ to the symmetry of $S^1\times S^2_\omega$, the $SO(3)$ symmetry of non-rotating sphere is explicitly broken to $U(1)_\phi$, more precisely for the $O(N)$ model on $S^1\times S^2$ introduction of $\omega$ leads to
\begin{equation}\label{eq:symbrk}
    U(1)_\tau\times SO(3)\times O(N) \underset{\omega\neq 0}{\longrightarrow} U(1)_\tau \times U(1)_\phi\times O(N)
\end{equation}
with discrete symmetry as $\theta\mapsto\pi-\theta$. The metric is $S^1\times S^2_\omega$ is written after coordinate transformation of $\phi \to\phi-i\omega\tau$ and this leads to identification \((\tau,\theta,\phi)\to(\tau+\beta,\theta,\phi)\), the metric is 
\begin{equation}
    ds^2_{S^1\times S^2_\omega} = d\tau^2+d\theta^2+\sin^2\theta(d\phi-i\omega d\tau)^2.
\end{equation}

We assume that the HS field $\sigma(x)$ respects the residual symmetries and therefore introduces no additional symmetry breaking, hence on these grounds $\sigma(\tau,\theta,\phi) =\sigma(\theta)$. For a general latitude-dependent profile, however, the Effective action depends non-locally on the complete function \(\sigma(\theta)\), and the gap equation cannot be evaluated efficiently by the constant-mass spectral decomposition discussed in \cite{David:2024pir}. 

\subsection{Fluid description and High-Temperature expansion}\label{ss:fl}

We consider a fluid approach to the Effective action \eqref{eq:eff-action} inspired from the geometry, the compactification $(\tau,\theta,\phi)\to(\tau+\beta,\theta,\phi-i\omega\beta)$ can be generated on the $\mathbb R_\tau \times S^2$ through a Killing Vector such that the integral curves are the compactification \cite{Jensen:2012jh}. Our method to use this method is inspired from \cite{Advant:2026cqt}.  We consider the Euclidean covering space $ \mathbb R_\tau\times S^2$
\begin{equation}\label{eq:eucmetr}
    ds^2_{\mathbb R\times S^2}=d\tau^2+d\theta^2+\sin^2\theta\,d\phi^2 .
\end{equation}
The complex Euclidean vector
\begin{equation}
    v=\beta\left(\partial_\tau-i\omega\partial_\phi\right),
\end{equation}
with norm $v^2=\beta^2\left(1-\omega^2\sin^2\theta\right)$ is  Killing as both $\partial_\tau$ and $\partial_\phi$ are Killing vectors of the
covering-space metric, one has $\nabla_{(\mu}v_{\nu)}=0.$ We restrict to $|\omega|<1$, so that $v^2$ is nowhere vanishing. The integral curves of $v$ are defined by
\begin{equation}\label{eq:vecdef}
    \frac{dV^\mu(s)}{ds}=v^\mu(V(s)),\qquad0\leq s\leq1 ,
\end{equation}
the integral curves $V^\mu$ are
\begin{equation}
    \tau(s)=\tau_0+\beta s,
    \qquad
    \theta(s)=\theta_0,
    \qquad
    \phi(s)=\phi_0-i\beta\omega s
\end{equation}
they verify the claim that the $v^\mu$ generates the identifications. Henceforth, we can write $S^1\times S^2_\omega$ as the quotient
\begin{equation}
    S^1\times S^2_\omega=\left(\mathbb R_\tau\times S^2\right)
/\langle e^v\rangle .
\end{equation}

The Killing vector $v^\mu$ can be interpreted as the fluid velocity (after analytical continuation and proper normalization) in Lorentzian spacetime. With $v^\mu$ in hand we can develop the Effective action in the language of Weyl invariant calculus developed in \cite{Loganayagam:2008is}. For a brief review see \cite{Advant:2026cqt} and for the construction specified to our problem, see \S\ref{as:sphflui}.

The vector $v^\mu$ is not normalized using the metric as Weyl transformation changes the proper length of the thermal circle but leaves its orbits and coordinate periodicity unchanged. We therefore assign \(v^\mu\) Weyl weight zero. With the convention that a quantity \(\Phi\) of Weyl weight \(w_\Phi\) transforms as
\(\Phi\rightarrow e^{w_\Phi\Omega}\Phi\), we have
\begin{equation}
g_{\mu\nu}\longrightarrow e^{2\Omega}g_{\mu\nu},
\; v^\mu\longrightarrow v^\mu \implies v_\mu\longrightarrow e^{2\Omega}v_\mu,
\;v^2\equiv g_{\mu\nu}v^\mu v^\nu\longrightarrow e^{2\Omega}v^2 .
\end{equation}

The local temperature and normalized thermal direction therefore transform
as
\begin{equation}
T\equiv\frac{1}{\sqrt{v^2}}\longrightarrow e^{-\Omega}T,
\;u^\mu\equiv Tv^\mu\longrightarrow e^{-\Omega}u^\mu,
\;u_\mu\longrightarrow e^{\Omega}u_\mu,
\end{equation}
so that \(u^\mu u_\mu=1\) remains invariant.

The Weyl transformation of the Hubbard--Stratonovich field follows from
the covariance of the conformal scalar operator
\begin{equation}
\mathcal O_\sigma=-\nabla^2+\zeta_c R+\sigma,
\end{equation}
under a Weyl transformation, conformal covariance requires $\sigma\to e^{-2\Omega}\sigma.$
More explicitly,
\begin{equation}
\mathcal O_{\sigma'}[e^{2\Omega}g]
\left(e^{-\frac{d-2}{2}\Omega}\phi\right) =e^{-\frac{d+2}{2}\Omega}
\mathcal O_\sigma[g]\phi,\quad \sigma'=e^{-2\Omega}\sigma .
\end{equation}
It follows that the saddle $\chi\equiv v^2\sigma$ is Weyl invariant.

The corresponding Weyl connection is
\begin{equation}
\mathcal A_\mu\equiv\frac12\partial_\mu\log v^2=\partial_\mu\log\sqrt{v^2}.
\end{equation}
Since \(v^2\rightarrow e^{2\Omega}v^2\), it transforms as $\mathcal A_\mu\to
\mathcal A_\mu+\partial_\mu\Omega$. For a scalar of Weyl weight \(w\), our convention for the Weyl-covariant derivative is
\begin{equation}
\mathcal D_\mu\Phi=\left(\nabla_\mu-w\mathcal A_\mu\right)\Phi,
\end{equation}
 this derivative converts the conformal Killing equation into $\mathcal D_{(\mu}v_{\nu)}=0.$
Finally, Weyl invariant measure is
\begin{equation}
d^3x\frac{\sqrt g}{(v^2)^{3/2}}, 
\end{equation}
hence we can write Effective action as
\begin{equation}\label{eq:lfde}
    \Gamma(\sigma) =\int d^3x\frac{\sqrt g}{(v^2)^{3/2}} \; \mathcal{I}(\chi).
\end{equation}
The $\mathcal{I}$ is a local Weyl invariant  scalar density constructed entirely from the  \(\chi\) and Weyl-covariant derivatives and curvature tensors. However $\Gamma(\sigma)$ as defined in \eqref{eq:eff-action} is a non-local quantity depending on $\sigma$ we can think of \eqref{eq:lfde} as a local high-temperature EFT description of it.

We expand $\mathcal{I}(\chi)$ in orders of derivatives and this expansion is controlled at high temperature because every projected derivative occurs in the dimensionless combination \(\widehat{\mathcal D}_{\mu}=\sqrt{v^{2}}\,P_{\mu}{}^{\nu}\mathcal D_{\nu}\), so that each derivative is accompanied by the local thermal length \(\sqrt{v^{2}}\propto\beta\). On the sphere, \(\beta\ll1\) therefore gives \(s,X,Q_{\chi}=O(\beta^{2})\) and the four-derivative invariants \(s^{2},sX,X^{2},Y=O(\beta^{4})\), while the dimensionless mass \(\chi=v^{2}\sigma\) is kept fixed and treated exactly. The successive derivative orders contribute to the Effective action as
\begin{equation}\label{eq:efact}
    \Gamma(\sigma) =\int d^3x\frac{\sqrt g}{(v^2)^{3/2}}\left[\mathcal{I}_0 +\mathcal{I}_2+\mathcal{I}_4 +\cdots \right]
\end{equation}
which identifies the derivative expansion directly with the high-temperature expansion in even powers of \(\beta\). The appearance of only even powers of \(\beta\) can be understood from the parity-even derivative(in Weyl invariants) expansion of the scalar determinant: on a smooth background without boundary, the independent local Weyl invariant scalars occur at even derivative order, integrations by parts. A complementary discussion using coordinates on $S^1\times S^2_\omega$ is given in \cite{Mauro:2026zus}.

\subsection{\texorpdfstring{Evaluation on $S^1\times S^2$}
{Evaluation on S1 x S2}}
\label{eq:s1s2}

We constructed the Weyl-covariant Effective action \eqref{eq:efact} in \S\ref{as:sphflui}. The calculation naturally separates into three parts. The background dependence is encoded entirely in the Weyl-invariant scalars \(s,X,Y,\ldots\), whose values on \(S^1\times S^2\) are given in \eqref{eq:anodef}.
We then determine the saddle perturbatively in the
derivative expansion. We have
\begin{equation}
    \Gamma(\sigma) =\int d^3x \frac{\sqrt{g}}{(v^2)^{3/2}} \mathcal I = \frac{2\pi}{\beta^2} \int_{-1}^1\,\frac{dx}{(1-\omega^2+\omega^2x^2)^{3/2}} \,\mathcal{I}.
\end{equation}

We expand the dimensionless saddle as
\begin{equation}\label{eq:xdex}
\chi =\chi_0+\chi_2+\chi_4+\mathcal O(\partial^6).
\end{equation}
In this approximation, we solve the gap equation order by order. Hence, we have
\begin{itemize}
    \item \textbf{Zero Derivatives}:
    At zero-derivative order, the off-shell Effective action is
    \begin{equation}
        \Gamma_{(0)}(\sigma) = \frac{2\pi}{\beta^2} \int_{-1}^1\,\frac{dx}{(1-\omega^2+\omega^2x^2)^{3/2}} \,\mathcal{I}_{0} = \frac{2\pi}{\beta^2} \int_{-1}^1\,\frac{dx}{(1-\omega^2+\omega^2x^2)^{3/2}} \,F(\chi).
    \end{equation}
    \begin{equation}
        \frac{\delta \Gamma_{(0)}(\sigma)}{\delta \sigma(\theta)} = 0\implies F'(\chi) = 0.
    \end{equation}
    We denote by $\bar{\chi}_0$ the solution of the gap equation $\left(F'(\bar \chi_0)=0\right)$, and define the on-shell Effective action as
    \begin{equation}
        \Gamma^{\rm on-shell}_{(0)} = \frac{4\pi}{\beta^2(1-\omega^2)} \mathcal{I}_{(0)}(\bar \chi_0).
    \end{equation}

    From the matching of coefficients with the determinant $\operatorname{Tr}\ln \mathcal{O}_\sigma$ (for details, see \S\ref{a:hksph}), we have
    \begin{equation}
        F(\chi) = -\frac{1}{2\pi}\left[\operatorname{Li}_3(e^{-\sqrt{\chi}})+\sqrt{\chi}\operatorname{Li}_2(e^{-\sqrt{\chi}})+\frac{\chi^{3/2}}{6}\right].
    \end{equation}
    The gap equation implies that $\bar\chi_0 =4\log^2\varphi_g$. Consequently, the zero-derivative/High-temperature HS saddle is
    \begin{equation}
        \bar\sigma_{(0)} = \frac{4\log^2\varphi_g}{\beta^2(1-\omega^2\sin^2\theta)}.
    \end{equation}

    Finally,
    \begin{equation}
        \Gamma^{\rm on-shell}_{(0)} = - \frac{8\zeta(3)}{5\beta^2(1-\omega^2)}.
    \end{equation}

    \item \textbf{Two Derivatives}:
    Up to two derivatives, the contribution is
    \begin{equation}
        \Gamma_{(2)} = \frac{2\pi}{\beta^2} \int_{-1}^1\,\frac{dx}{(1-\omega^2+\omega^2x^2)^{3/2}} \,\mathcal I_{(2)},
    \end{equation}
    with $\mathcal{I}_{(2)} =F_s(\chi)s+F_X(\chi)X+K(\chi)Q_\chi$. Here, the coefficients $F_s(\chi),F_X(\chi),K(\chi)$ are determined by matching with the actual determinant. Since we are solving for \eqref{eq:xdex} hence at four derivatives order $Q_\chi$ is evaluated at $\chi_0$ and $K_\chi$ term doesn't contribute. From the matching in \S\ref{a:hksph}, we have
    \begin{equation}\label{eq:Fss}
        \begin{split}
            &F_X(\chi) =-\frac16F'(\chi),\qquad F_s(\chi) =-\frac13F'(\chi)-\frac{\sqrt{\chi}}{24\pi}\coth\left(\frac{\sqrt{\chi}}{2}\right),\\
        \end{split}
    \end{equation}

    For the expansion \eqref{eq:xdex}, the gap equation at second order is
    \begin{equation}
        F''(\bar\chi_0)\chi_2+ F'_s(\bar\chi_0)s+F'_X(\bar\chi_0)X =0
        \implies
        \chi_2(x)=\frac16 X(x)+\frac{4\log\varphi_g}{3\sqrt{5}}s(x).
    \end{equation}

    At two-derivative order, $Q_\chi$ does not contribute, and the on-shell action is
    \begin{equation}
        \Gamma^{\rm on-shell}_{(2)} =\frac{\sqrt{5}\log\varphi_g}{9}\frac{\omega^2}{1-\omega^2}.
    \end{equation}

    \item \textbf{Four Derivatives}:

    The four-derivative contribution contains a fairly large number of invariant scalars, but the four-derivative scalars containing explicit derivatives of $\chi$ vanish when evaluated at $\bar\chi_0$. Therefore, the four-derivative contribution to the on-shell density is
    \begin{equation}
        \mathcal I^{\rm on-shell}_{(4)} = F_{ss}(\bar\chi_0)s^2+F_{sX}(\bar\chi_0)sX+F_{XX}(\bar\chi_0)X^2+F_Y(\bar\chi_0)Y -\frac{\big(F'_s(\bar\chi_0)s+F'_X(\bar\chi_0)X\big)^2}{2F''(\bar\chi_0)}.
    \end{equation}

    Using the definitions of $s,X,Y$, we obtain
    \begin{equation}
        \begin{split}
            \Gamma^{\rm on-shell}_{(4)} &= \frac{2\pi\beta^2}{1-\omega^2}\Bigg[\frac{2\omega^4}{5}C_{ss}-\frac{\omega^2(1-2\omega^2)}{3}C_{sX}\\
                      &\qquad\qquad\qquad+\frac{8\omega^4-4\omega^2+3}{6} C_{XX}-\frac{4\omega^2(5-\omega^2)}{15}C_Y\Bigg].
        \end{split}
    \end{equation}
    Here, the $C$'s are undetermined on-shell combination of  Wilson coefficients of the theory. They are
    \begin{equation}
        \begin{split}
            C_{ss}&\equiv F_{ss}(\bar\chi_0)- \frac{ F_s'(\bar\chi_0)^2 }{2F''(\bar\chi_0)},\qquad
            C_{sX}\equiv F_{sX}(\bar\chi_0)-\frac{ F_s'(\bar\chi_0)F_X'(\bar\chi_0)}{F''(\bar\chi_0)},\\
            C_{XX}&\equiv F_{XX}(\bar\chi_0)-\frac{F_X'(\bar\chi_0)^2}{ 2F''(\bar\chi_0)},\qquad
            C_Y \equiv F_Y(\bar\chi_0).
        \end{split}
    \end{equation}
\end{itemize}

The partition function of the theory defined by \eqref{eq:largeN} is
\begin{equation}\label{eq:logZht}
    \begin{split}
         \log Z(\beta,\omega)={}&\frac{8N\zeta(3)}{5\beta^2(1-\omega^2)}-\frac{N\sqrt5\log\varphi_g}{9}
        \frac{\omega^2}{1-\omega^2}\\
        &-\frac{2\pi N\beta^2}{1-\omega^2}
        \Bigg[\frac{2\omega^4}{5}C_{ss}-\frac{\omega^2(1-2\omega^2)}{3}C_{sX}\\
        &\hspace{20mm}
            +\frac{8\omega^4-4\omega^2+3}{6}C_{XX}-\frac{4\omega^2(5-\omega^2)}{15}C_Y\Bigg]+O(\beta^4).
    \end{split}
\end{equation}

The coefficients $C$s are not uniquely determinable form the hydrostatic description only, we can only determine a family of them as four derivative contribution is invariant under translation with respect to some parameter $\bar \rho$
\begin{equation}
    (C_{ss},C_{sX},C_{XX},C_{Y}) \mapsto  (C_{ss}+6\bar\rho,C_{sX}-4\bar\rho,C_{XX},C_{Y}+\bar\rho).
\end{equation}
The above translation ambiguity can be understand as the total derivatives ambiguity discussed in \cite{Advant:2026cqt}.

We evaluate these coefficients on the sphere with the heat kernel method, we have sketched the solution in \S\ref{a:hksph}, the coefficients we get are 
\begin{equation}\label{eq:wl66}
    \begin{split}
        C_{ss} &= \frac{1}{\pi} \left( \frac{17}{720} - \frac{\sqrt{5}}{320\log\varphi_g} - \frac{2\sqrt{5}\log\varphi_g}{144} \right),\;\, C_{XX} = -\frac{\sqrt{5}}{720\pi \log\varphi_g} , \\
C_{sX} &= \frac{1}{\pi} \left( \frac{\sqrt{5}}{240\log\varphi_g} - \frac{1}{45} \right), \;\, C_{Y} = \frac{1}{\pi} \left( \frac{1}{240} - \frac{\sqrt{5}}{960\log\varphi_g} \right).
    \end{split}
\end{equation}

So finally we have the partition function on the sphere as
\begin{equation}\label{eq:lozspfull}
    \begin{split}
         \log Z(\beta,\omega)={}&\frac{2\pi N}{1-\omega^2}\,\Biggl[\frac{4\zeta(3)}{5\pi\beta^2}-\frac{\sqrt5\log\varphi_g}{18\pi}\omega^2+\frac{\beta^2}{10800\pi}
        \Bigg[\frac{15\sqrt{5}}{2\log\varphi_g}-20\left(1+\frac{\sqrt{5}}{2\log\varphi_g}\right)\omega^2\\
        &\qquad\qquad+\left(46+\frac{13\sqrt{5}}{2\log\varphi_g}+60\sqrt{5}\log\varphi_g\right)\omega^4\Bigg]+ O(\beta^4)\Biggr].
    \end{split}
\end{equation}
From \eqref{eq:lozspfull} we observe that all positive powers of $\beta$ are accompanied with a polynomial (in $\omega^2$) of same degree, this is the form already predicted in \cite{Advant:2026cqt}.  
The first two Wilson coefficients (coefficients of $\beta^{-2}$ and $\beta^0$ terms)
were already obtained by the existing methods (see \cite{Mauro:2026zus}), the other four Wilson coefficients \eqref{eq:wl66} obtained here are new to this paper.

 In $\omega =0$ limit we have
\begin{equation}
    \log Z_{(\omega =0)} =\frac{4\pi N}{\beta^2}\left[\frac{2\zeta(3)}{5\pi}+\frac{\sqrt{5}}{2880\pi\log\varphi_g}\beta^4+O(\beta^6)\right],
\end{equation} 
this matches exactly to the Eq.~(3.20) of \cite{David:2024pir} (with $r=1$), which is partition function of at non-rotating sphere and functions on the first consistency check of our calculation.

In the small-$\omega$ limit ($\omega\to0$), the first two orders in the derivative expansion through $O(\omega^2)$ are
\begin{equation}\label{eq:logzsav}
    \log Z_{\omega\to0} =\frac{8N\zeta(3)}{5\beta^2}(1+\omega^2) - \frac{N\sqrt{5}\log\varphi_g}{9} \omega^2 +  O(\omega^4,\beta^2).
\end{equation}

\eqref{eq:logzsav} matches exactly with the small-$\omega$ corrections presented in Eq.~(3.12) of \cite{Mauro:2026zus}, comparison can be made more explicitly by analytically continuing $\omega$ used here as $\omega\to -i\omega$ and explicitly keeping $r$ as radius of sphere in \eqref{eq:logzsav}, we simply set $r=1$ in this article. 

\section{Near the Hot Equator}\label{sec:hote}

The semi-universal limit is well understood in terms of the large local Temperature, as the local temperature defined as
\begin{equation}
    T^2=\frac{1}{\beta^2(1-\omega^2\sin^2\theta)},
\end{equation}
putting this in the language adopted in the \cite{Anand:2025mfh}, in $d=4$ we have the $\gamma_4(\theta)$
blowing up in $\omega\to1$ limit, the story is little different in $d=3$ as $\gamma_3(\theta)$ blow up in $\omega \to 1$ limit at and near the equator only, as the $\gamma_3(\theta) =\sqrt{T}$ we call the semi-universal limit in $d=3$ as Hot-Equator\footnote{We thank Sridip Pal for the terminology of \emph{Hot-Equator}.}. This limit is then elegantly captured by \cite{Komargodski:2026ain} with the following scalings in metric of $S^1\times S^2_\omega$\,
\begin{equation}\label{eq:rdfesc}
    \varepsilon \equiv1-\omega^2,\quad \theta=\frac{\pi}{2}+\sqrt{\varepsilon}\,y,\quad \phi=\varepsilon u, \;\; u\sim u+\frac{2\pi}{\varepsilon}, 
\end{equation}
these specific choice of redefinition can be understood as near the equator $(\delta\theta=\theta-\pi/2)$ and 
\begin{equation}
    1-\omega^2\sin^2\theta =\varepsilon +(\delta\theta)^2 + O(\varepsilon(\delta\theta)^2,(\delta\theta)^4)
\end{equation}
and this implies that $\delta\theta\sim\sqrt{\varepsilon}$ and in order to make the $d\tau d\phi$ $O(\varepsilon)$ we have $\phi=\varepsilon u$.
In these metric redefinitions and limit $\varepsilon\to0$ the leading order metric is
\begin{equation}
    ds^2_{\varepsilon\to0} =\varepsilon (dy^2-2idud\tau+(1+y^2)d\tau^2) + O(\varepsilon^2),
\end{equation}
from this we take the $\varepsilon$ at front out a Weyl factor,and then we define the pp wave metric as
\begin{equation}\label{eq:ppmet}
    g^{\rm pp}_{\mu\nu}\equiv \lim_{\varepsilon\to0}\frac{1}{\varepsilon}g_{\mu\nu},\, \implies ds^2_{\rm pp} =(1+y^2)d\tau^2 +dy^2-2id\tau du.
\end{equation}
In the same Weyl frame the operator $\mathcal O_\sigma$ also transforms as, 
\begin{equation}
    \begin{split}
        \mathcal{O}_\sigma =-\nabla^2_g +\frac18 R[g] +\sigma(\theta) =\frac{1}{\varepsilon}\left( -\nabla^2_{g_{\rm pp}}+\frac18 R[g_{\rm pp}] +\varepsilon\sigma(\theta)\right),
    \end{split}
\end{equation}
so the correct operator in the same Weyl frame is \(\mathcal{O}_\sigma \, \varepsilon = \bar{\mathcal{O}}\) with the expansion of HS field $\sigma(\theta)$ is 
\begin{equation}
    m(y) \equiv \lim_{\varepsilon\to 0}\,\varepsilon \,\sigma\left(\frac{\pi}{2}+\sqrt{\varepsilon}\,y\right),\qquad \bar{\mathcal O} =-\nabla^2_{g_{\rm pp}} +m(y),
\end{equation}
because $R[g_{\rm pp}]=0$. The Effective action on the pp-wave geometry is
\begin{equation}\label{eq:effacpp}
    \Gamma(m) =\frac12\operatorname{Tr}\log \left(-\nabla^2+m(y)\right) +\frac{\mu_0^2}{\lambda_0}\int_{\rm pp} m(y),
\end{equation}
here we have the same critical prescription as on the full sphere.

Similarly to \S\ref{ss:fl} we can formulate the same local high-temperature EFT by considering the covering space of \eqref{eq:ppmet} with the unwrapped $\tau$ coordinate and $v^\mu_{\rm pp}=\beta\partial_\tau$ as the Killing Vector such that the integral curve generates the identification $(\tau,u,y)\to(\tau+\beta,u,y)$, upto two derivatives we compute the Effective action and partition function on the pp-wave is
\begin{equation}\label{eq:lzpp}
    \log Z_{\rm pp} =\frac{2\pi N}{\varepsilon}\left[\frac{4\zeta(3)}{5\pi\beta^2}-\frac{\sqrt5\log\varphi_g}{18\pi}+O(\beta^2)\right],
\end{equation}
The limiting fluid variables, the pp-wave values of the Weyl-covariant
invariants, and their explicit contribution to the Effective action are provided in \S\ref{sub:coeff}.

The way \cite{Komargodski:2026ain} showed that the $\log Z_{\rm pp}$ is the correct limit of the partition function on the $S^1\times S^2$ in $\omega\to1$ limit, so we also verify the same here, \eqref{eq:logZht} in the $\omega\to1$ limit is 
\begin{equation}\label{eq:lnZsmemi}
\begin{split}
    \log Z(\beta,\omega)_{\omega\to1} ={}& \frac{2\pi N}{1-\omega^2} \Bigg[ \frac{4\zeta(3)}{5\pi\beta^2} - \frac{\sqrt{5}\log\varphi_g}{18\pi} + O(\beta^2)\Bigg].\\
\end{split}
\end{equation}
This is the structure of the partition function proposed by \cite{Anand:2025mfh}, and it has also the predicted simple pole structure and its semi-universal residue, also that the \eqref{eq:lzpp} matches with the \eqref{eq:lnZsmemi}. Thus the theory on the pp-wave computes precisely the semi-universal residue of the $\omega\to1$ pole.

\section{Semi-universal Residue}\label{sec:surmai}

The preceding analysis reduces the near light-like rotation ($\omega \to1$) limit of the sphere partition function to evaluation on the pp-wave geometry. The pole structure $\varepsilon^{-1}$ appears due the regulated volume of the pp-wave spacetime \cite{Komargodski:2026ain}. In the context of the \cite{Anand:2025mfh} we have 
\begin{equation}
    \log Z_{\rm pp} = \frac{h(\beta)}{\varepsilon} =\frac{h(\beta)}{1-\omega^2},
\end{equation}
here $h(\beta)$ is the theory dependent (semi-universal) residue of $\log Z_{\rm pp}$ and here it is computed for $O(N)$ model. 

\subsection{Worldline Heat Kernel and Effective action}
The Effective action \eqref{eq:effacpp} has the $\operatorname{Tr}\log \bar{\mathcal O}$, we represent this determinant in Schwinger parametrization \cite{Schwinger:1951nm} via Schwinger proper time $t$ as
\begin{equation}\label{eq:strep}
    \operatorname{Tr}\log \bar{\mathcal O} =- \int_0^\infty\frac{dt}{t}\, \operatorname{Tr} e^{-t\bar{\mathcal O}}. 
\end{equation}

The $\operatorname{Tr}e^{-t\bar{\mathcal O}}$ can be written in form of the trace of coincident Heat Kernel on the pp-wave space as
\begin{equation}
   \operatorname{Tr} e^{-t\bar{\mathcal O}} =\operatorname{Tr} K_{\rm pp}(x;t),
\end{equation}
here $x$ is coordinates on pp-wave spacetime, \cite{Giombi:2008vd} the heat kernel on the  pp-wave spacetime with thermal identification $\tau \to \tau +\beta$ can be written as sum of all images of the heat kernel in a covering space of pp-wave spacetime with unwrapped $\tau$ and the image is defined as map
\begin{equation}
    f^n : (\tau,u,y)\mapsto (\tau+n\beta,u,y)
\end{equation}
and consequently 
\begin{equation}
    K_{\rm pp}(x;t) =\sum_{n\in \mathbb Z} K_{\texttt{pp}}(x,f^nx;t).
\end{equation}

For the \eqref{eq:strep} we have
\begin{equation}
    \operatorname{Tr}\log \bar{\mathcal O} =- \int_0^\infty\frac{dt}{t}\, \int d^3x\sqrt{g_{\rm pp}}\;K_{\rm pp}(x;t) =- \sum_{n\in\mathbb Z}\int_0^\infty\frac{dt}{t}\, \int d^3x\sqrt{g_{\rm pp}}\;K_{\texttt{pp}}(x,f^nx;t),
\end{equation}

The free heat kernel on the $\texttt{pp}$-wave spacetime using the De-Witt iterative procedure \cite{Vassilevich:2003xt} is
\begin{equation}
    K_{\texttt{pp}}(x,f^nx';t) =\frac{1}{(4\pi t)^{3/2}}\, \Delta^{1/2}_{\rm vvm} \,\exp\left\{-\frac{\Sigma(x,f^nx')}{2t}\right\} \Xi(x,f^nx',t),
\end{equation}
here $\Delta_{\rm vvm}$ is Van Vleck–Morette determinant and $\Sigma$ is Synge's world function, for an free scalar Laplacian considered here heat kernel is exact hence $\Xi =1$. For the free heat kernel in coincident limit we have
\begin{equation}\label{eq:datamet}
\begin{split}
    \Delta^{1/2}_{\rm vvm}(x,f^nx) =\sqrt{\frac{n\beta}{\sinh(n\beta)}},\quad \Sigma(x,f^nx) = \frac{n^2\beta^2}{2} +n\beta y^2\tanh\left(\frac{n\beta}{2}\right),
\end{split}
\end{equation}
this result is for the $n$-th image. This gives the free heat kernel in the $\texttt{pp}$-wave spacetime for the $n$-th image is
\begin{equation}\label{eq:kpfren}
    K^{\rm free}_{\texttt{pp}}(x,f^nx;t)  =\frac{1}{(4\pi t)^{3/2}}\sqrt{\frac{n\beta}{\sinh(n\beta)}}\,\exp\left(-\frac{n^2\beta^2}{4t} -\frac{n\beta y^2}{2t}\tanh\left(\frac{n\beta}{2}\right)\right)
\end{equation}

Heat Kernel for operator $\bar{\mathcal O}$ can be represented in worldline path integral representation \cite{Fliegner:1994zc}, for the $n$-th image the worldline representation is 
\begin{equation}\label{eq:wlrep}
    K_{\texttt{pp}}(x,f^nx;t) = \int_{X(0)=X}^{X(t)=f^n(X)} \mathcal D X \;\exp\left[-\int_0^td\lambda \left(\frac14 g^{\rm pp}_{\mu\nu}\dot X^\mu(\lambda) \dot X^\nu(\lambda)+ m(X(\lambda))\right)\right],
\end{equation}
here $\lambda$ is a worldline parameter and  inspired from free heat kernel that the kinetic part of \eqref{eq:wlrep} reduces to the \eqref{eq:kpfren}, 
\begin{equation}\label{eq:wlavs}
    K_{\texttt{pp}}(x,f^nx;t) =  K^{\rm free}_{\texttt{pp}}(x,f^nx;t) \left\langle\exp\left[-\int_0^t d\lambda \; m\left(X(\lambda)\right)\right]\right\rangle,
\end{equation}
here the worldline average is defined as
\begin{equation}\label{eq:wlavgr}
    \langle \mathcal{J} \rangle = \frac{ \int\limits_{X(0) = X}^{X(t) = f^n(X)} \mathcal{D}X \, \Theta(X) \, \mathcal{J}(X)}{ \int\limits_{X(0) = X}^{X(t) = f^n(X)} \mathcal{D}X \, \Theta(X)},
\end{equation}
here $\Theta(X)$ is the kinetic part of \eqref{eq:wlrep}. Finally we present a formula for $\log\det$ as
\begin{equation}\label{eq:ldetwl}
     \operatorname{Tr}\log \bar{\mathcal O} =- \sum_{n\in \mathbb Z}\int_0^\infty\frac{dt}{t}\,\int d^3x\sqrt{g_{\rm pp}}\,  \;K^{\rm free}_{\texttt{pp}}(x,f^nx;t) \left\langle\exp\left[-\int_0^t d\lambda \; m\left(X(\lambda)\right)\right]\right\rangle. 
\end{equation}

\subsection{High-Temperature asymptotics}

In order to compute the high-temperature limit of the partition function we would approximate the worldine average in \eqref{eq:ldetwl}, to get this high-$ T$ limit we approximate the \eqref{eq:ldetwl} in $\beta\to0$ limit. 

We let $t=\beta^2\rho$ and $\widetilde{m}(y)=\beta^2 m(y)$, we define
\begin{equation}\label{eq:tnm}
    \mathcal{T}_n[m] =\left\langle \exp\left[-t\int_0^1 dz\,m(y(z))\right]\right\rangle =\left\langle \exp\left[-\rho \int_0^1 dz\,\widetilde m(y(z))\right]\right\rangle,
\end{equation}
here in $\mathcal{T}_n[m]$ we take a worldline geodesic and expand around the geodesic in the $\beta\to0$ limit the corrections due to curvature are generated by the fluctuations around the geodesic. We have done that expansion  as prescribed in \cite{Fliegner:1994zc} in \S\ref{app:pphk}, the procedure is to write the propagator for fluctuations and then we sum the diagrams with constraints from Wick's theorem. The approximate value of the \eqref{eq:tnm} is achieved by summing over all the connected Feynman graphs ($\log \mathcal{T}_n[m]=$ sum of all connected graphs) and is given by
\begin{equation}
\mathcal{T}_n[m]= e^{-\rho \widetilde m(y)} \exp\big[\beta^2 l_2+\beta^4l_4+\beta^6l_6\cdots\big],
\end{equation}
functions $l_2,l_4,l_6$ are first few corrections and are listed in \eqref{eq:l2}-\eqref{eq:l6}. 

In the $\beta\to 0$ limit the \eqref{eq:kpfren} has also an expansion in $\beta^2$ as
\begin{equation}
    \sqrt{\frac{n\beta}{\sinh(n\beta)}}\,\exp\left(-\frac{n^2\beta^2}{4t} -\frac{n\beta y^2}{2t}\tanh\left(\frac{n\beta}{2}\right)\right) = e^{-\frac{n^2(1+y^2)}{4\rho}} Q_n(y,\rho;\beta),
\end{equation}
here $Q_n=1+\beta^2q_2+\beta^4q_4+\cdots$ we can write the first few $q$s as
\begin{equation}\label{eq:qgeo}
    \begin{split}
         & q_2 = -\frac{n^2}{12} + \frac{n^4y^2}{48\rho}, \quad 
    q_4 = \frac{n^4}{160} -\frac{11n^6y^2}{2880\rho} + \frac{n^8y^4}{4608\rho^2},\\
    &q_6 = -\frac{61n^6}{120960} + \frac{83n^8y^2}{161280\rho} -\frac{17n^{10}y^4}{276480\rho^2} + \frac{n^{12}y^6}{663552\rho^3}.
    \end{split}
\end{equation}

Finally, 
\begin{equation}\label{eq:hkppow}
     K_{\texttt{pp}}(x,f^nx;t) =\frac{e^{-n^2(1+y^2)/4\rho -\rho \widetilde m(y)} }{(4\pi \beta^2\rho)^{3/2}} \sum_{\alpha=0}^\infty \beta^{2\alpha}B_{2\alpha}[y,n,\rho;\widetilde m]
\end{equation}
and
\begin{equation}\label{eq:trlog}
   \begin{split}
        \operatorname{Tr}\log \bar{\mathcal O} &=- \sum_{n\in\mathbb Z}\int_{0}^\infty\frac{dt}{t}\, \int d^3x\sqrt{g_{\rm pp}}\;K_{\texttt{pp}}(x,f^nx;t)\\
        & = - \frac{1}{4\varepsilon \beta^2 \sqrt{\pi} }\sum_{n\in\mathbb Z}\int_{-\infty}^\infty dy\,\int_0^\infty d\rho\, \rho^{-\frac52} \,e^{-n^2(1+y^2)/4\rho -\rho\widetilde m(y)}\sum_{\alpha=0}^\infty \beta^{2\alpha}B_{2\alpha}[y,n,\rho;\widetilde m],
   \end{split}
\end{equation}
we let that $B_{2\alpha}$ is a finite Laurent polynomial in $\rho$ and an even polynomial in $n$, at every fixed order \(\alpha\), only finitely many coefficients \(b_{2\alpha;p,r}\) are nonzero, hence 
\begin{equation}\label{eq:someB}
    B_{2\alpha} = \sum_{p\in\mathbb Z} \sum_{r\geq 0} b_{2\alpha;p,r}(\widetilde m;y) \rho^p n^{2r}.
\end{equation}
In these conventions the proper-time integral can be easily evaluated as
\begin{equation}\label{eq:ptint}
    I^{(n)}_p[\widetilde m ;y] \equiv \int_0^\infty d\rho \rho^{\,p-\frac52} e^{-n^2(1+y^2)/4\rho -\rho\widetilde m(y)} 
\end{equation}
the regulated integral for $n=0$ and exact integral for $n\neq 0$ is
\begin{equation}\label{eq:ptint0}
    I^{0}_p = \Gamma_{\rm E}\left(p-\frac32\right) \widetilde m(y)^{3/2-p},\;\;\; I^{(n)}_p = 2\left(\frac{n^2(1+y^2)}{4\widetilde m(y)}\right)^{\frac{p}{2}-\frac34} K_{p-3/2}\left(|n|\sqrt{(1+y^2)\widetilde m(y)}\right),
\end{equation} 
here $\Gamma_E$ Euler's gamma function and $K$ is modified Bessel function of second kind. The regularization of $I^0_p$ provides a compact implementation of the
subtraction of the power divergences in the zero-winding sector.
Its finite part is fixed by the same zero-temperature critical
renormalization conditions used in the original Effective action.
No subtraction is required for the nonzero-winding sectors.  So we have
\begin{equation}\label{eq:}
    \Gamma(m) = -\frac{1}{8\sqrt{\pi}\varepsilon\beta^2} \sum_{\iota =0}^\infty \beta^{2\iota} \mathcal{G}_{2\iota}[\widetilde m]
\end{equation}

\begin{equation}
    \mathcal{G}_{2\iota} = \int_{-\infty}^\infty dy\left[\sum_{p\in\mathbb Z} b_{2\iota;p,o}I^0_p +2\sum_{n=1}^\infty\sum_{p\in \mathbb Z}\sum_{r\geq 0} b_{2\iota;,p,r} n^{2r} I^{(n)}_p\right]
\end{equation}

Using half-integer representation of modified Bessel function of second kind ($K$) we can do the $n$ sum in terms of poly-logarithms as
\begin{equation}
    K_{N_p+1/2}(x) = \sqrt{\frac{\pi}{2x}} e^{-x}\sum_{j=0}^{N_p}\frac{(N_p+j)!}{j!(N_p-j)!(2x)^j}, \quad N_p = \begin{cases} 
p - 2, & p \ge 2, \\ 
1 - p, & p \le 1. 
\end{cases}
\end{equation}

and 
\begin{equation} 
    \begin{split}
        \mathcal{J}_{p,r}[y;\widetilde m] &= 2\sum_{n=1}^\infty I_p^{(n)} n^{2r} \\
     &= 4\left(\frac{\chi(y)}{4\widetilde m(y)^2}\right)^{\frac{p}{2}-\frac34} \sqrt{\frac{\pi}{2\sqrt{\chi(y)}}} \sum_{j=0}^{N_p}\frac{(N_p+j)!}{j!(N_p-j)!(2\sqrt{\chi})^j } \;\operatorname{Li}_{2-p-2r+j}(e^{-\sqrt{\chi(y)}}).
    \end{split}
\end{equation}

Hence, 
\begin{equation}
    \mathcal{G}_{2\iota} = \int_{-\infty}^\infty dy\left[\sum_{p\in\mathbb Z} b_{2\iota;p,o}I^0_p +\sum_{p\in \mathbb Z}\sum_{r\geq 0} b_{2\iota;,p,r}\,\mathcal{J}_{p,r} \right]
\end{equation}

Now we have a off-shell Effective action \eqref{eq:effacpp} in a high-$T$ expansion and now we solve the gap equation with a high-$T$ ansatz such that $\widetilde m_*(y)$ solves the gap equation
\begin{equation}
  \widetilde  m_*(y) = \widetilde m_0(y) +\beta^2 \widetilde m_2(y) +\beta^4 \widetilde m_4(y) +\cdots .
\end{equation}

The functional derivative and the $k$-th functional derivative at $\widetilde m_0$ as
\begin{equation}
    \mathcal G^{(k)}_{2\iota}[\widetilde m_0] \equiv \frac{\delta^{(k)}\mathcal G_{2\iota}}{\delta \widetilde m(y_1) \cdots\delta \widetilde m(y_k)} \Bigg|_{\widetilde m =\widetilde m_0},
\end{equation}
so the gap equation order by order is
\begin{equation}
    \mathcal{G}^{(1)}_0[\widetilde m_0](y) =0 ,\;\; \int_{-\infty}^\infty dy_1 \mathcal{G}^{(2)}_0[\widetilde m_0](y,y_1) \widetilde m_2(y_1) + \mathcal{G}_2^{(1)}[\widetilde m_0](y) =0,
\end{equation}

Using this method we can in principle collect coefficients upto the arbitrary large orders in $\beta^2$, in this analysis we report our results upto $\beta^4$.
The partition function is
\begin{equation}\label{eq:logZppf}
    \log Z_{\rm pp} =\frac{2\pi N}{\beta^2(1-\omega^2)} \left[\sum_{i=0}^{\infty} \beta^{2i}\;\Gamma^{\rm on-shell}_{2i}\right] .
\end{equation}

The algorithm to get the $\Gamma^{\rm on-shell}_{2i}$ is pretty straightforward as one can get the $B_{2\alpha}$-s following the steps of \S\ref{app:pphk} consequently getting $b$-s and $I$-s and then applying the gap equation, the one numerical process is the $y$ integral here, other all can be done by symbolic recursions. The order $\beta^2$ and $\beta^4$ we computed numerically.
\begin{equation}\label{eq:finallogZpp}
    \begin{split}
    \log Z_{{\rm pp}}=\frac{2\pi N}{(1-\omega^2)}\Bigg[&\frac{4\zeta(3)}{5\pi\beta^2}-\frac{\sqrt5\log\varphi}{18\pi}+0.003216943326\,\beta^2\\
    &-0.0024762242577\,\beta^4+O(\beta^6)\Bigg].
\end{split}
\end{equation}

The coefficient of  $\beta^2$ in \eqref{eq:lozspfull} agrees to displayed numerical value with coefficient of $\beta^2$ in \eqref{eq:finallogZpp}  as in $\omega\to1$ limit coefficient from \eqref{eq:lozspfull} is
\begin{equation}
    \frac{13 + \frac{2\sqrt{5}}{\log\varphi_g} + 30\sqrt{5}\log\varphi_g}{5400\pi} \approx 0.003216943326,
\end{equation}
and this matching further strengthens  
\begin{equation}
    (1-\omega^2)\log Z_{\rm pp} = \lim_{\omega\to 1}\,(1-\omega^2)\log Z_{S^1\times S^2}.
\end{equation}

\subsection{Low-temperature asymptotics}

The low-temperature expansion is an expansion of the finite-temperature saddle about the zero-temperature critical saddle \(m(y)=0\). At zero temperature there are no thermal excitations, and critical tuning makes the renormalized vacuum functional stationary at \(m=0\). At large but finite \(\beta\), the dilute thermal population produces an exponentially small source for \(m(y)\). The finite-temperature profile is obtained by applying the response kernel of the critical vacuum to this source.

For fixed positive null momentum \(k\), the pp-wave Hamiltonian in the background \(m(y)\) is \cite{Komargodski:2026ain}
\begin{equation}\label{eq:hpp}
H_k[m]=-\frac{1}{2k}\frac{ d^2}{ dy^2}
+\frac{k}{2}(1+y^2)+\frac{m(y)}{2k}.
\end{equation}
At \(m=0\), it is a harmonic oscillator of unit frequency, with spectrum
\begin{equation}\label{eq:free-pp-spectrum}
E^{(0)}_{a,k}=a+\frac12+\frac{k}{2},
\qquad a\in\mathbb N_0,\quad k>0.
\end{equation}
Here \(a\) labels the  oscillator levels.

We separate the renormalized Effective action into its vacuum and thermal parts,
\begin{equation}\label{eq:thermal-split-definition}
\Gamma_{\beta,{\rm ren}}[m]=\Gamma_{\rm{vac,ren}}[m]+\Gamma_{\rm{th},\beta}[m],
\qquad
\Gamma_{{\rm th},\beta}[m]\longrightarrow0
\quad\text{as}\quad\beta\longrightarrow\infty.
\end{equation}
For an extensive Effective action, \(\Gamma_{\rm{vac,ren}}\) is the renormalized zero-winding contribution. Equivalently, it may be obtained from the zero-temperature limit of the Effective-action density.

Let $G_0=\left(-\nabla_{{\rm pp}}^2\right)^{-1}$,
the difference between the vacuum determinants at \(m\neq0\) and \(m=0\) is
\begin{equation}
\Gamma_{\rm{vac,ren}}[m]-\Gamma_{\rm{vac,ren}}[0]
=\frac12\operatorname{Tr}_0\log(1+G_0m)
+\Gamma_{\rm{ct}}[m]-\Gamma_{\rm{ct}}[0].
\end{equation}
Critical tuning imposes
\begin{equation}\label{eq:finetun}
\left.
\frac{\delta\Gamma_{\rm{vac,ren}}[m]}
{\delta m(y)}\right|_{m=0}=0.
\end{equation}
The linear counterterm therefore cancels the vacuum tadpole
\(\frac12\operatorname{Tr}_0(G_0m)\), and hence
\begin{equation}\label{eq:vacsubb}
\Gamma_{\rm{vac,ren}}[m]-\Gamma_{\rm{vac,ren}}[0]
= \frac12\operatorname{Tr}_0 \left[\log(1+G_0m)-G_0m\right].
\end{equation}
Choosing \(\Gamma_{\rm{vac,ren}}[0]=0\), the complete renormalized functional becomes
\begin{equation}\label{eq:vac-th-split}
\Gamma_{\beta,{\rm ren}}[m]
=\frac12\operatorname{Tr}_0
\left[\log(1+G_0m)-G_0m\right]
+\Gamma_{\rm{th},\beta}[m].
\end{equation}

It is useful to display separately the contribution already present at \(m=0\):
\begin{equation}\label{eq:source-response-split}
\Gamma_{\beta,\rm{ren}}[m]
=\Gamma_{\rm{th},\beta}[0]
+\frac12\operatorname{Tr}_0
\left[\log(1+G_0m)-G_0m\right]
+\Gamma_{\rm{th},\beta}[m]-\Gamma_{\rm{th},\beta}[0].
\end{equation}
 The first is the free contribution of the dilute thermal gas. The second describes the response of the zero-temperature critical vacuum to \(m(y)\). The final term describes how the thermal free energy changes when the saddle is displaced away from \(m=0\).

We denote the partition function as
\begin{equation}
    \log Z_{\rm pp} =\frac{2\pi N}{\varepsilon} W(\beta),
\end{equation}
in this convention we have, $\Gamma_{\beta,\rm ren}=\beta f_\beta$ and $W(\beta) =-\beta f_\beta$
\begin{equation}
  W_0(\beta) =\frac{1}{2\pi\beta}\sum_{n\geq1} \frac{1}{n^2\sinh(n\beta/2)},\,\; W_0(\beta) = \frac{1}{\pi \beta}\left(q+\frac{q^2}{4}+\frac{10q^3}{9}+\cdots\right).
\end{equation}

The lower threshold of \eqref{eq:free-pp-spectrum} is \(E_0=1/2\), reached for \(a=0\) and \(k\rightarrow0^+\). The exponential sectors are therefore organized by $q\equiv e^{-\beta/2}$.

Moreover, the factor \(e^{-\beta k/2}\) localizes the null momentum at \(k\sim\frac{1}{\beta}\),
since a harmonic-oscillator wavefunction at fixed \(k\) has width
\(\Delta y\sim k^{-1/2}\), the thermally occupied states extend over
\(\Delta y\sim\sqrt{\beta}\).

To derive the thermal source, we work with the free-energy density
\(f_{{\rm th},\beta}=\Gamma_{{\rm th},\beta}/\beta\). Its spectral representation is
\begin{equation}\label{eq:thermal-free-energy}
f_{{\rm th},\beta}[m]
=\frac{1}{\beta}\int_0^\infty\frac{ dk}{2\pi}
\sum_{a=0}^\infty
\log\left(1-e^{-\beta E_{a,k}[m]}\right).
\end{equation}
At \(m=0\), this gives the free dilute-gas contribution. The same thermal population also generates the source
\begin{equation}\label{eq:thermal-source-definition}
J_\beta(y)\equiv
\left.
\frac{\delta f_{{\rm th},\beta}[m]}
{\delta m(y)}
\right|_{m=0}.
\end{equation}
Using the Hellmann--Feynman relation
\begin{equation}
\left.
\frac{\delta E_{a,k}[m]}{\delta m(y)}
\right|_{m=0}
=\frac{|\psi_{a,k}(y)|^2}{2k},
\end{equation}
we obtain
\begin{equation}\label{eq:thermal-source-spectral}
J_\beta(y)
=\int_0^\infty\frac{ dk}{4\pi k}
\sum_{a=0}^\infty
n_{ B}\left(E^{(0)}_{a,k}\right)
|\psi_{a,k}(y)|^2,
\qquad
n_{ B}(E)=\frac{1}{e^{\beta E}-1}.
\end{equation}
and performing the oscillator sum with the Mehler kernel gives
\begin{equation}\label{eq:thermal-source-exact}
J_\beta(y)
=\frac{1}{4\pi}\sum_{n=1}^\infty
\frac{1}{
\sqrt{\sinh(n\beta)
\left[n\beta+2y^2\tanh(n\beta/2)\right]}}.
\end{equation}
At large \(\beta\), the $n=1$ dominates. With \(y=\sqrt{\beta}\,\eta\),
\begin{equation}\label{eq:thsl}
J_\beta(\sqrt{\beta}\eta)=\frac{q}{\sqrt{\beta}}j_1(\eta)+ \frac{q^2}{\sqrt{\beta}}j_2(\eta)+O\left(\frac{q^3}{\sqrt{\beta}}\right),\quad j_1(\eta) =\frac{1}{4\pi\sqrt{\eta^2+\frac12}},\; j_2(\eta)=\frac{1}{4\pi\sqrt{\eta^2+1}}
\end{equation}

 \eqref{eq:thsl} is the perturbation that moves the system away from the critical vacuum. The response is supplied by the quadratic term in the zero-temperature determinant \(H_0\), the Effective action begins as
\begin{equation}\label{eq:lsd}
f_\beta[m]
=f_{{th},\beta}[0]
+f^{(2)}_{\rm vac}
+\langle J_\beta,m\rangle+f^{(3)}_{\rm vac} +f^{(2)}_{\rm th} +\cdots
\end{equation}

For profile $m(y)$, the vacuum contribution follows from the zero-winding sector of \eqref{eq:nker}--\eqref{eq:c0}. At zero-winding ($n=0$) at fixed $\beta$, the classical bridge becomes $\bar y(t)=y$, the free kernel is $(4\pi s)^{-3/2}$, and the  covariance reduces to $2s[\min(t,t')-tt']$, where $s$ is the proper time and $t,t'\in[0,1]$. The potential-dependent kernel therefore factorizes as
\begin{equation}
    K_{\mathrm{pp},\,n=0}^{[m]}(x,x;s)
=\frac{1}{4\pi s}\,
K_{-\partial_y^2+m(y)}(y,y;s).
\end{equation}
Since $(4\pi s)^{-1}=\int \frac{d^2q}{(2\pi)^2}e^{-sq^2}$, the vacuum determinant per unit longitudinal coordinate volume coincides with the flat three-dimensional determinant for a potential depending only on $y$. Expanding about $m=0$ thus gives the massless bubble and triangle integrals with external momenta along $y$, yielding \eqref{eq:f2vac}--\eqref{eq:f3vac}. The linear term is removed by the critical subtraction in \eqref{eq:vacsubb}.

\begin{equation}\label{eq:f2vac}
    f^{(2)}_{\rm vac}[m]=-\frac14\operatorname{Tr}(G_0mG_0m) =-\frac{1}{32}\int\frac{dp}{2\pi} \frac{|\tilde m(p)|^2}{|p|} =\frac12 \langle m,H_0m\rangle ,\;\; H_0=-\frac{1}{16|p|}
\end{equation}
\begin{equation}\label{eq:f3vac}
    f^{(3)}_{\rm vac}[m]=\frac16\operatorname{Tr}(G_0mG_0mG_0m)=\frac{1}{48}\int \frac{dp_1 dp_2}{(2\pi)^2} \;\frac{\tilde m(p_1)\tilde m(p_2) \tilde m(p_3)}{|p_1p_2p_3|},\quad p_3=-p_1-p_2
\end{equation}
in these we have computed the loop-integrals with help of Feynman Parametrization. Let the $m=m_1+m_2$ and putting on saddle condition we have
\begin{equation}\label{eq:m1m2}
    H_0m_1+J_1 =0,\quad \langle H_0m_1+J_1,m_2\rangle=0,
\end{equation}

First solve for the $m_1$ and let $m_1=\frac{q}{\beta}S_1(\eta)$, then we have 
\begin{equation}
    \tilde S_1(s) = 16|s|\tilde j_1(s) =\frac{8}{\pi}|s|K_0\left(\frac{|s|}{\sqrt{2}}\right),
\end{equation}
here the convention for Fourier transform is $\tilde F(s) =\int d\eta e^{-is\eta}F(\eta)$ and $s=\sqrt{\beta}\,p$. We have $S_1(\eta)$ from the inverse transform of $\tilde S_1(s)$ and we have
\begin{equation}\label{eq;s1eta}
    S_1(\eta)=\frac{8}{\pi^2}\left[\frac{1}{R_\eta^2}-\frac{\eta}{R_\eta^3}\sinh^{-1}(\sqrt{2}\eta)\right],\quad R_\eta =\sqrt{\eta^2+\frac12}.
\end{equation}

\eqref{eq;s1eta} gives us the first correction to the saddle away from vacuum. The first correction to the Effective action is
\begin{equation}\label{eq:fbere}
    \begin{split} 
        f_{\beta} -f_{\rm th,\beta}[0]&=\frac12 \langle J_1,H_0^{-1}J_1\rangle -\langle J_1,H_0^{-1}J_1\rangle \\
        &=\frac{8q^2}{\beta}\int_{-\infty}^\infty \frac{ds}{2\pi}|s||\tilde j_1(s)|^2 =\frac{2q^2}{\pi^3\beta},
    \end{split}
\end{equation}
this gives us the first correction to the $W(\beta)$ at $O(q^2)$ is $-2/\pi^3$.

At $O(q^3)$ we can use $m_1$ to write the hamiltonian in scaled coordinates $\eta$ and $\kappa =k\beta$, and then we can use a calculation similar to problems in linear response theory and corrections to it. The leading order gives us the same answer as \eqref{eq:fbere}, the corrections provide us with the response of saddle at quadratic order. Finally all the corrections to $O(q^3)$ are
the free gas, quadratic response of $m_1$ and its coupling to $m_2$ and the cubic vertex vacuum term $(f^{(3)}_{\rm vac})$.

We complete the calculation that we sketched above in \S\ref{ap:ld}. So at the order of $q^3$ we have $W(\beta)$ as
\begin{equation}
    W(\beta) = \frac{q}{\pi\beta}+\left(\frac{1}{4\pi\beta}-\frac{2}{\pi^3}\right) q^2 +\left(\frac{10}{9\pi\beta}+\frac{4\beta}{3\pi^3}-\frac{12\log 2}{\pi^3}+\frac{28\zeta(3)}{\pi^5}\right) q^3 + O(q^4),
\end{equation}
moreover, we have the partition function in low temperature approximation is
\begin{equation}
    \begin{split}
        \log Z_{{pp}} &= \frac{2\pi N}{1-\omega^2} \left[ \frac{e^{-\beta/2}}{\pi\beta} + \left(\frac{1}{4\pi\beta} - \frac{2}{\pi^3}\right)e^{-\beta} \right. \\
        &\qquad\qquad \left. + \left(\frac{10}{9\pi\beta} + \frac{4\beta-36\log 2}{3\pi^3} + \frac{28\zeta(3)}{\pi^5}\right) e^{-3\beta/2} + O(e^{-2\beta}) \right]
    \end{split}
\end{equation}

Throughout this expansion, powers of $q=e^{-\beta/2}$ label exponential sectors whose coefficients may contain powers or logarithms of $\beta$. The remainder notation denotes omitted sectors beginning at $q^4$. 

\section{Discussion}\label{sec:diss}

In this work, we studied the leading large-$N$ thermal partition
function of the critical $O(N)$ model on the rotating background
$S^1\times S^2$. At high temperature and fixed $|\omega|<1$, the
partition function is described by a Weyl-covariant hydrostatic
derivative expansion, which we computed through four derivatives.
In the near-light-speed limit at fixed $\beta$, the leading contribution
comes from the region near the equator and is described by the same
theory on a pp-wave background. The agreement, to numerical precision,
of the $\beta^2$ coefficient obtained from the two calculations gives a
nontrivial check of our results.

An important feature of the rotating problem is that the
HS saddle is not spatially constant. In the
small-angular-velocity analysis of \cite{Mauro:2026zus}, the correction
to the saddle does not contribute to the optimized Effective action at
the order considered there, so that the constant saddle is sufficient
for that calculation. Our finite-$\omega$ result reproduces their
coefficient in the small-$\omega$ limit. At finite $\omega$, however,
the proper length of the thermal circle depends on the latitude, and
the leading saddle develops a corresponding $\theta$ dependence fixed
by the local Tolman temperature. This dependence becomes important as
$\omega\to1$.

The near-light-speed limit realizes the semi-universal behavior
discussed in \cite{Anand:2025mfh,Mukherjee:2026dfu},
\[
\log Z(\beta,\omega)
=
\frac{h(\beta)}{1-\omega^2}
\left(1+O(1-\omega)\right).
\]
The pole is geometric, while the residue $h(\beta)$ depends on the
theory. In the pp-wave description, the pole comes from the divergent
effective volume near the equator, while $h(\beta)$ is determined by
the interacting theory on the limiting geometry.

At high temperature, $h(\beta)$ has an expansion in even powers of
$\beta$, with coefficients determined by the worldline expansion and
the self-consistent saddle. Our low-temperature analysis gives an asymptotic expansion in the sectors \(q=e^{-\beta/2}\). The integrals used are finite, although we do not establish uniform control of the expansion for simultaneous \(k\to0\) and large \(|y|\). A numerical solution of the full pp-wave gap equation would be useful to study the global finite-temperature saddle beyond this expansion.

It would be interesting to determine $h(\beta)$ at intermediate
temperature by solving the pp-wave gap equation numerically, thereby
connecting the high- and low-temperature regimes. At $O(N^0)$, one
should also include fluctuations of the Hubbard--Stratonovich field.
It would further be useful to determine the terms finite as
$\omega\to1$, which should contain information about the geometry away
from the equator. The same pp-wave method (in particular the Worldline expansion) may also be generalised and  applied to other
interacting large-$N$ CFTs as Gross--Neveu model (see \cite{Diatlyk:2023msc,Chamati:2011zz}) and Chern--Simons mattter theories (see \cite{Aharony:2012ns,Jain:2013py}.

\acknowledgments
I am grateful to Shiraz Minwalla for several fruitful discussions throughout this project and to the authors of \cite{Advant:2026cqt} for early access to their manuscript, which helped the hydrostatic EFT formulation used here. I thank Jyotirmoy Mukherjee and Shiraz Minwalla for useful comments on the draft.
I acknowledge the use of ChatGPT GPT 5.6-Sol (OpenAI) for assistance with organizing notes, occasional rephrasing, and LaTeX formatting. I independently developed and verified all scientific and mathematical content and take full responsibility for the final manuscript.
This work is supported by the Department of Atomic Energy, Government of India, under Project Identification Number RTI-4012. I also acknowledge my debt to the people of India for their continued support of basic scientific research.

\appendix
\section{Weyl-covariant fluid dictionary}\label{ap:wifd}
For the Fluid description, we introduce the local temperature and the
normalized thermal direction
\begin{equation}
T=\frac{1}{\sqrt{v^{2}}},
\qquad
u^{\mu}=\frac{v^{\mu}}{\sqrt{v^{2}}}=T v^{\mu},
\qquad
u^{\mu}u_{\mu}=1.
\end{equation}
Here \(v^\mu\) is assumed to be nowhere null, \(v^{2}\neq0\). The transverse
projector is
\begin{equation}
P_{\mu\nu}
=
g_{\mu\nu}-u_{\mu}u_{\nu}
=
g_{\mu\nu}-\frac{v_{\mu}v_{\nu}}{v^{2}},
\qquad
P_{\mu\nu}u^{\nu}=0.
\end{equation}
The expansion, acceleration, shear and vorticity associated with \(u^\mu\)
are
\begin{equation}
\Theta=\nabla_{\mu}u^{\mu},
\qquad
a_{\mu}=u^{\nu}\nabla_{\nu}u_{\mu},
\end{equation}
\begin{equation}
\Sigma_{\mu\nu}
=
P_{\mu}{}^{\alpha}P_{\nu}{}^{\rho}
\left(
\nabla_{(\alpha}u_{\rho)}
-\frac{\Theta}{d-1}P_{\alpha\rho}
\right),
\qquad
\omega^{(u)}_{\mu\nu}
=
P_{\mu}{}^{\alpha}P_{\nu}{}^{\rho}
\nabla_{[\alpha}u_{\rho]}.
\end{equation}
In \(d=3\), the coefficient of \(P_{\alpha\rho}\) in the shear tensor is
\(\Theta/2\). For the stationary rotating configuration,
\begin{equation}
\Theta=0,
\qquad
\Sigma_{\mu\nu}=0,
\end{equation}
whereas \(a_\mu\) and \(\omega^{(u)}_{\mu\nu}\) can be nonzero.

More generally, let \(v^\mu\) be a nowhere-null conformal Killing vector,
\begin{equation}
\nabla_{(\mu}v_{\nu)}
=
\psi,g_{\mu\nu},
\qquad
\psi=\frac{1}{d}\nabla_{\mu}v^{\mu}.
\end{equation}
Contracting the conformal Killing equation with \(v^\mu v^\nu\) gives
\begin{equation}
v^{\mu}\nabla_{\mu}v^{2}
=
2\psi v^{2},
\qquad
v^{\mu}\nabla_{\mu}\log v^{2}
=
2\psi.
\end{equation}

Under a Weyl transformation, the thermal vector is held fixed,
\begin{equation}
g_{\mu\nu}\longrightarrow e^{2\Omega}g_{\mu\nu},
\qquad
v^{\mu}\longrightarrow v^{\mu}.
\end{equation}
Consequently,
\begin{equation}
v_{\mu}\longrightarrow e^{2\Omega}v_{\mu},
\qquad
v^{2}\longrightarrow e^{2\Omega}v^{2},
\qquad
T\longrightarrow e^{-\Omega}T,
\qquad
u^{\mu}\longrightarrow e^{-\Omega}u^{\mu}.
\end{equation}
The mixed projector \(P_{\mu}{}^{\nu}\) is Weyl invariant, while
\begin{equation}
P_{\mu\nu}\longrightarrow e^{2\Omega}P_{\mu\nu},
\qquad
P^{\mu\nu}\longrightarrow e^{-2\Omega}P^{\mu\nu}.
\end{equation}

The Weyl connection naturally associated with \(v^\mu\) is
\begin{equation}
\mathcal A_{\mu}
=
\frac{1}{2}\partial_{\mu}\log v^{2}.
\end{equation}
Since
\begin{equation}
v^{\mu}\mathcal A_{\mu}=\psi,
\end{equation}
and \(v^{2}\rightarrow e^{2\Omega}v^{2}\), the connection transforms as
\begin{equation}
\mathcal A_{\mu}
\longrightarrow
\mathcal A_{\mu}+\partial_{\mu}\Omega.
\end{equation}
For a scalar \(\Phi\) of Weyl weight \(w\), we use
\begin{equation}
\mathcal D_{\mu}\Phi
=
\left(\nabla_{\mu}-w\mathcal A_{\mu}\right)\Phi,
\end{equation}
together with the corresponding connection terms for tensor indices. In
particular, \(v_\nu\) has Weyl weight two and
\begin{equation}
\mathcal D_{\mu}v_{\nu}
=
\nabla_{\mu}v_{\nu}
-\mathcal A_{\mu}v_{\nu}
+\mathcal A_{\nu}v_{\mu}
-g_{\mu\nu}\mathcal A_{\rho}v^{\rho}.
\end{equation}
The conformal Killing equation then becomes
\begin{equation}
\mathcal D_{(\mu}v_{\nu)}
=
\nabla_{(\mu}v_{\nu)}
-g_{\mu\nu}\mathcal A_{\rho}v^{\rho}
=0.
\end{equation}

The projected dimensionless derivative is defined by
\begin{equation}
\widehat{\mathcal D}_{\mu}
\equiv
\sqrt{v^{2}}P_{\mu}{}^{\nu}\mathcal D_{\nu}.
\end{equation}
The Weyl-covariant vorticity of the thermal vector is
\begin{equation}
\varpi_{\mu\nu}
\equiv
2\mathcal D_{[\mu}v_{\nu]}.
\end{equation}
Since \(\mathcal D_{(\mu}v_{\nu)}=0\), this can equivalently be written as
\begin{equation}
\varpi_{\mu\nu}=2\mathcal D_{\mu}v_{\nu}.
\end{equation}
We adopt the convention
\begin{equation}
\operatorname{tr}(\varpi^{2})
\equiv
\varpi_{\mu}{}^{\nu}\varpi_{\nu}{}^{\mu}
=-\varpi_{\mu\nu}\varpi^{\mu\nu},
\quad s\equiv-\frac18\operatorname{tr}(\varpi^{2}).
\end{equation}

In three dimensions, the ordinary Schouten tensor is
\begin{equation}
\mathsf P_{\mu\nu}
\equiv
R_{\mu\nu}-\frac14Rg_{\mu\nu}.
\end{equation}
The corresponding Weyl-covariant Schouten tensor and its trace are
\begin{equation}\label{eq:a22}
    \mathcal S_{\mu\nu}
\equiv
\mathsf P_{\mu\nu}
+\nabla_{\mu}\mathcal A_{\nu}
+\mathcal A_{\mu}\mathcal A_{\nu}
-\frac12\mathcal A^{2}g_{\mu\nu},
\
\mathcal S
\equiv
g^{\mu\nu}\mathcal S_{\mu\nu}.
\end{equation}
The remaining geometric building blocks required through four derivatives
are
\begin{equation}
X\equiv v^{2}\mathcal S,
\qquad
B_{\mu}\equiv\mathcal S_{\mu\nu}v^{\nu},
\qquad
B_{\mu}^{\perp}\equiv P_{\mu}{}^{\nu}B_{\nu},
\qquad
Y\equiv v^{2}B_{\mu}^{\perp}B_{\perp}^{\mu}.
\end{equation}
With the curvature conventions used above, these quantities satisfy the
three-dimensional identity
\begin{equation}
v^{\mu}B_{\mu}=2s-X.
\end{equation}

The parity-even purely geometric scalars needed through fourth derivative
order may therefore be chosen as
\begin{equation}
\mathcal O(\partial^{2}):
\qquad
s,\quad X,
\end{equation}
and
\begin{equation}
\mathcal O(\partial^{4}):
\qquad
s^{2},\quad sX,\quad X^{2},\quad Y.
\end{equation}

For a space-dependent \(\chi\), the two-derivative scalar is
\begin{equation}
Q_{\chi}
\equiv
\widehat{\mathcal D}_{\mu}\chi,
\widehat{\mathcal D}^{\mu}\chi
=
v^{2}P^{\mu\nu}
\mathcal D^{\mu}\chi\mathcal D_{\nu}\chi.
\end{equation}
A term proportional to
\(F(\chi)\widehat\Delta\chi\) can be exchanged for a \(Q_\chi\) term by
integration by parts, so it need not be included independently at second
order. At fourth order, it is convenient to define
\begin{align}\label{eq:defs}
\widehat\Delta\chi
\equiv
\widehat{\mathcal D}_{\mu}
\widehat{\mathcal D}^{\mu}\chi,
\,
H_{\mu\nu}
\equiv
\widehat{\mathcal D}_{(\mu}
\widehat{\mathcal D}_{\nu)}\chi,
\,
J_{s}
\equiv
\widehat{\mathcal D}_{\mu}\chi
\widehat{\mathcal D}^{\mu}s,
\,
J_{X}
\equiv
\widehat{\mathcal D}_{\mu}\chi
\widehat{\mathcal D}^{\mu}X.
\end{align}
A convenient parity-even set containing derivatives of \(\chi\)
at fourth order is
\begin{equation}
sQ_{\chi},\;
XQ_{\chi},\;
Q_{\chi}^{2},\;
J_{s},\;
J_{X},\;
Q_{\chi}\widehat\Delta\chi,\;
(\widehat\Delta\chi)^{2},\;
H_{\mu\nu}H^{\mu\nu}.
\end{equation}
This set is understood modulo integration by parts, Bianchi identities and
three-dimensional tensor identities.

\subsection{On the Sphere}\label{as:sphflui}

All these quantities (relevant for upto four derivative evaluation) can be explicitly computed on the manifold $\mathbb R_\tau \times S^2$, we have the Weyl connection has the only nonzero component
\begin{equation}
    \mathcal A_{\theta}=\frac12\partial_{\theta}\log f=-\frac{\omega^{2}\sin\theta\cos\theta}{f(\theta)}.
\end{equation}

With $f=v^2/\beta^2$, $x=\cos\theta$.
For reference, the nonzero components of the ordinary Schouten tensor are
\begin{equation}
    \mathsf P_{\tau\tau}=-\frac12,
    \qquad
    \mathsf P_{\theta\theta}=\frac12,
    \qquad
    \mathsf P_{\phi\phi}=\frac12\sin^{2}\theta.
\end{equation}
Writing $A(\theta)\equiv\mathcal A_{\theta}$, the Schouten
tensor is diagonal and has components
\begin{align}
    \mathcal S_{\tau\tau}
      =-\frac12\left(1+A^{2}\right),\quad
    \mathcal S_{\theta\theta}
      =\frac12+\partial_{\theta}A+\frac12A^{2},\quad
    \mathcal S_{\phi\phi}
      =\frac12\sin^{2}\theta
        +\sin\theta\cos\theta\,A
        -\frac12\sin^{2}\theta\,A^{2}.
\end{align}
The four purely geometric fourth-order invariants are obtained simply as
\begin{equation}
    s(x)^{2},\quad s(x)X(x),\quad X(x)^{2},\quad Y(x),
\end{equation}
and  independent scalar invariants reduce to
\begin{equation}\label{eq:anodef}
    \begin{split}
         s(x)
      &=-\beta^{2}\frac{\omega^{2}x^{2}}{f(x)},\\
    X(x)
      &=\frac{\beta^{2}}{2f(x)}
        \left[1-\omega^{4}+\omega^{4}x^{2}
        -4\omega^{2}x^{2}\right],\\
    Y(x)
      &=-\beta^{4}
        \frac{\omega^{2}(1-\omega^{2})^{2}(1-x^{2})}
        {f(x)^{2}}.
    \end{split}
\end{equation}

The Cotton tensor vanishes on $\mathbb R_\tau\times S^{2}$, so it supplies no
additional invariant on this background. 
\begin{enumerate}
    \item \textbf{At zero derivatives}
    At zero derivatives only available scalar is $\chi=v^2\sigma$, so we have $\mathcal{I}_0 =F(\chi)$. 
    \item \textbf{At two derivatives}
    At two derivatives, the independent parity-even invariants are
       \begin{equation}
           s,\qquad X,\qquad Q_\chi \equiv\widehat{\mathcal D}_\mu\chi\,\widehat{\mathcal D}^{\mu}\chi.
       \end{equation}
     The most general local density is therefore
      \begin{align}
      \mathcal I_{(2)}&=F_s(\chi)s+F_X(\chi)X+K(\chi)Q_\chi.
      \end{align}
    \item \textbf{At four derivatives}
    The background-only algebraic parity-even invariants relevant to the
on-shell action at four-derivative order are
\begin{equation}
    s^2,\qquad sX,\qquad X^2,\qquad Y.
\end{equation}
The corresponding contribution to the density is
\begin{equation}
    \mathcal I_{(4)}
    = F_{ss}(\chi)s^2+F_{sX}(\chi)sX+F_{XX}(\chi)X^2+F_Y(\chi)Y.
\end{equation}

Using  \eqref{eq:a22}-\eqref{eq:defs} contractions of Weyl-covariant Schouten tensor, we define
\begin{equation}
    \begin{split}
        &\mathcal S_{\mu\nu}^{\perp}
        \equiv P_\mu{}^\alpha P_\nu{}^\beta\mathcal S_{\alpha\beta},
        \qquad
        \mathcal S_{\langle\mu\nu\rangle}
        \equiv\mathcal S_{\mu\nu}^{\perp}
        -\frac12P_{\mu\nu}P^{\alpha\beta}\mathcal S_{\alpha\beta},\\
        &\mathcal I_{\mathcal S\chi\chi}
        \equiv(v^2)^2\mathcal S_{\langle\mu\nu\rangle}
        \mathcal D^\mu\chi\,\mathcal D^\nu\chi,\qquad
        \mathcal I_{\mathcal SH}
        \equiv(v^2)^2\mathcal S_{\langle\mu\nu\rangle}
        \mathcal D^\mu\mathcal D^\nu\chi.
    \end{split}
\end{equation}
A representative set of four-derivative operators containing
explicit derivatives of $\chi$, including mixed contractions of
derivatives of $\chi$ with the background tensors, is
\begin{equation}
    \begin{split}
        \mathcal I_{(4)}^{(\nabla\chi)}
        \supset{}&L_{sQ}(\chi)sQ_\chi+L_{XQ}(\chi)XQ_\chi
        +L_{QQ}(\chi)Q_\chi^2\\
        &+L_{Js}(\chi)J_s+L_{JX}(\chi)J_X
        +L_{Q\Delta}(\chi)Q_\chi\widehat\Delta\chi\\
        &+L_{\Delta\Delta}(\chi)(\widehat\Delta\chi)^2
        +L_{HH}(\chi)H_{\mu\nu}^{(\chi)}H_{(\chi)}^{\mu\nu}\\
        &+L_{\mathcal S\chi\chi}(\chi)\mathcal I_{\mathcal S\chi\chi}
        +L_{\mathcal SH}(\chi)\mathcal I_{\mathcal SH}.
    \end{split}
\end{equation}

\end{enumerate}

\subsection{In pp wave geometry}\label{ap:ppf}

We use the pp-wave metric and thermal vector
\begin{equation}\label{eq:ppg}
    ds_{\rm pp}^{2}=dy^{2}-2i\,d\tau\,du+h(y)d\tau^{2},
    \qquad h(y)=1+y^{2},\qquad v=\beta\partial_{\tau},
\end{equation}
where $\tau\sim\tau+\beta$ and $y\in\mathbb R$.
The local temperature, normalized thermal direction, and
Weyl-invariant saddle variable are
\begin{equation}\label{eq:ppv}
    v^{2}=\beta^{2}h,\qquad T=\frac{1}{\beta\sqrt h},
    \qquad u^{\mu}=\frac{\delta^\mu_\tau}{\sqrt h},
    \qquad\chi=v^{2}m=\beta^{2}h\,m.
\end{equation}
The Weyl connection and acceleration are
\begin{equation}\label{eq:ppa}
    \mathcal A=\frac{y}{h}\,dy,
    \qquad
    a=-\frac{y}{h}\,dy,
    \qquad
    \Theta=0,
    \qquad
    \Sigma_{\mu\nu}=0.
\end{equation}
The only independent nonzero component of the thermal vorticity is
$\varpi_{yu}=2i\beta y/h$.
Consequently, the two-derivative invariants evaluate to
\begin{equation}\label{eq:ppi}
    s=-\beta^{2}\frac{y^{2}}{h},
    \qquad
    X=\frac{\beta^{2}}{2h}(2-3y^{2}),
    \qquad
    Q_{\chi}=\beta^{2}h\,(\partial_y\chi)^{2},
\end{equation}
and four-derivatives invariants are 
\begin{equation}
\begin{aligned}
\mathcal{O}(\partial^4):\qquad
\Big\{&
s^2,\; sX,\; X^2,\; Y,\;
sQ_\chi,\; XQ_\chi,\; Q_\chi^2,\;
J_s,\; J_X,\\
&Q_\chi \widehat{\Delta}\chi,\;
(\widehat{\Delta}\chi)^2,\;
H_{\mu\nu}H^{\mu\nu},\;
I_{S\chi\chi},\;
I_{SH}
\Big\}.
\end{aligned}
\end{equation}

Through two derivatives, the Effective action is
\begin{equation}\label{eq:ppact}
    \begin{split}
        \left.\Gamma[\chi]\right.=\frac{2\pi }{\beta^{2}\varepsilon}
        \int_{-\infty}^{\infty}\frac{dy}{h^{3/2}}\bigg[ F(\chi)-\frac{\beta^{2}y^{2}}{h}F_s(\chi)
            +\frac{\beta^{2}(2-3y^{2})}{2h}F_X(\chi)+\beta^{2}h\,K(\chi)(\partial_y\chi)^{2} \bigg].
    \end{split}
\end{equation}
The coefficient functions are the same as those appearing in the
sphere calculation; only the background values of the invariants
have changed.

\section{Local Wilson Coefficients}\label{a:hksph}

We already have $x=\cos\theta$, $f(x)=1-\omega^2+\omega^2x^2$ and $v^2=\beta^2f(x)$ we introduce $u=\tau/\beta$ and the co-rotating angle $\varphi=\phi+i\omega\tau$.  The Weyl-invariant metric and mass variable are
\begin{equation}\label{eq:thvar}
 \widehat g_{\mu\nu}=\frac{g_{\mu\nu}}{v^2},\;\chi=v^2\sigma,\;d\mu=\frac{d^3x\sqrt g}{(v^2)^{3/2}}.
\end{equation}
In these coordinates the metric takes the Kaluza--Klein form as
\begin{equation}
    d\hat{s}^2=(du+ a_idy^i)^2+ \hat h_{ij}dy^idy^j, \;\;\; a_\phi=-\frac{i\omega(1-x^2)}{\beta f(x)}
\end{equation}
the base metric is
\begin{equation}
    \hat h_{xx}=\frac{1}{\beta^2f(x)(1-x^2)},\;\; \hat h_{\phi\phi} =\frac{1-x^2}{\beta^2f(x)^2},\;\; \sqrt{\hat h}=\frac{1}{\beta^2f(x)^{3/2}}
\end{equation}
The connection $ a_i$ is the Kaluza--Klein gauge field, distinct from
the Weyl connection used in the fluid description.  We define 
\begin{equation}
 \mathcal F_{ij}=\partial_i a_j-\partial_j a_i =b\,\epsilon_{ij}, \;\;\epsilon_{ij}\epsilon^{ij}=2,\;\;R_{\hat h}=R[\hat h].
\end{equation}
we have
\begin{equation}\label{eq:kkgeo}
 b=\frac{2i\beta\omega x}{\sqrt f},\;\;
 R_{\hat h}=\frac{2\beta^2}{f}
 \left[1-\omega^4+(\omega^4-5\omega^2)x^2\right],\;\;
 \widehat R=R_{\hat h}-\frac{b^2}{2}.
\end{equation}

The large-$N$ determinant is that of the conformal scalar operator
\begin{equation}
 \mathcal O_\sigma=-\nabla^2+\frac{R}{8}+\sigma.
\end{equation}
Writing $k_n=2\pi n$, its $n$th thermal Fourier mode in the thermal frame is
\begin{equation}\label{eq:modeop}
 \mathcal O_n=-D_n^2+\chi+k_n^2+\mathcal V,
 \qquad
 D_{n,i}=\nabla_i^{(h)}-ik_n a_i,
 \qquad
 \mathcal V=\frac{R_h}{8}-\frac{b^2}{16}.
\end{equation}
Thus $[D_{n,i},D_{n,j}]=-ik_nb\epsilon_{ij}$ on scalars.  Here every $\beta$ dependence is in the local fluid variables as $s,X,Y$ and $\chi$.

The fluid invariants evaluated on this background satisfy
\begin{equation}\label{eq:dict}
 b^2=4s,\qquad R_{\hat h}=4X+2s,\qquad (\nabla_hb)^2=4Y.
\end{equation}
 In the hydrostatic counting
\begin{equation}
 b=O(\partial),\qquad R_{\hat h},\mathcal V=O(\partial^2),\qquad
 \chi,k_n=O(\partial^0).
 \label{eq:count}
\end{equation}
so the local action is
\begin{equation}\label{eq:hydexp}
    \begin{split}
       \Gamma&=\int_{S^1\times S^2}d\mu\,\Bigl[F(\chi)+F_s(\chi)s+F_X(\chi)X\\
             &+F_{ss}(\chi)s^2+F_{sX}(\chi)sX
             +F_{XX}(\chi)X^2+F_Y(\chi)Y+O(\partial^6)\Bigr].
    \end{split}
\end{equation}

\subsection{Computing the coefficients}
\label{sub:coeff}

For the off-shell functional, the Schwinger representation is
\begin{equation}
 \frac12\operatorname{Tr}\log\mathcal O
 =-\frac12\sum_{n\in\mathbb Z}
 \int_0^\infty\frac{dt}{t}\,
 \operatorname{Tr}_{h}e^{-t\mathcal O_n}.
 \label{eq:sch}
\end{equation}

The integrated Seeley--DeWitt expansion can now be read from the standard
coefficients from \cite{Vassilevich:2003xt}. We discard all total derivatives of $\chi$ in here for reasons we would mention later, this works as we are restricted to four derivative order only. 
Consequently, the terms of order $\partial^0$, $\partial^2$, and
$\partial^4$ come from the relevant parts of the usual $a_0$, $a_2$,
$a_4$, $a_6$, and $a_8$ coefficients. The cofficients $a_0,a_2,a_4,a_6$ can be directly read off from \cite{Vassilevich:2003xt} Eq~(4.26)-(4.28) and $a_8$ cofficient from \cite{Amsterdamski:1989bt}. 
Moreover we can just derive the relevant part from $a_8$ as we require only the term quartic in $b$. 
\begin{equation}
 \frac{1}{\operatorname{Vol}}\operatorname{Tr}e^{-t(-D_n^2)}=\frac{1}{4\pi t}\frac{t k_n b}{\sinh(t k_n b)}=\frac{1}{4\pi t}\left(1-\frac{t^2k_n^2b^2}{6}+\frac{7t^4k_n^4b^4}{360}+O(\partial^6)\right)
\end{equation}
In the present background their
sum is
\begin{equation}\label{eq:hkall}
    \begin{split}
     \operatorname{Tr}_{h}e^{-t\mathcal O_n}={}&\frac{e^{-t(\chi+k_n^2)}}{4\pi t}\int d^2x\sqrt h\,
 \Bigg\{1+t\left(\frac{R_h}{6}-\mathcal V\right)-t^2\frac{k_n^2b^2}{6}\\
 &\quad+t^2\left(\frac{R_h^2}{60}-\frac{R_h\mathcal V}{6}+\frac{\mathcal V^2}{2}\right)\\
 &\quad+t^3k_n^2\left(-\frac{R_hb^2}{20}+\frac{\mathcal Vb^2}{6}+\frac{(\nabla_hb)^2}{60}\right)
 +t^4\frac{7k_n^4b^4}{360}+O(\partial^6)\Bigg\}.
 \end{split}
\end{equation}
The first line of \eqref{eq:hkall} gives the zero- and two-derivative
terms.  The remaining two lines give the complete integrated contribution
at four derivatives. 

\paragraph{Renormalization.}
Before extracting the coefficient functions, we must renormalize the
zero-derivative determinant.  Introducing a proper-time cutoff, its bare
contribution is
\begin{equation}
 F_{\rm{det},\Lambda}(\chi)
 =-\frac{1}{8\pi}\int_{1/\Lambda^2}^{\infty}
 \frac{dt}{t^2}\,e^{-t\chi}
 \sum_{n\in\mathbb Z}e^{-tk_n^2}.
 \label{eq:wbare}
\end{equation}
The ultraviolet structure is most transparent after Poisson resummation,
\begin{equation}
 \sum_{n\in\mathbb Z}e^{-4\pi^2n^2t}
 =\frac{1}{\sqrt{4\pi t}}
  \sum_{\ell\in\mathbb Z}e^{-\ell^2/(4t)}.
 \label{eq:pois}
\end{equation}
Every nonzero winding sector is exponentially suppressed as $t\to0$.
Hence the entire $\chi$-dependent ultraviolet divergence comes from the
zero-winding term.  Returning momentarily to the original three-dimensional
variables, this contribution is
\begin{align}
 \Gamma_{\rm{det},\Lambda}^{(l=0)}
 &=-\frac{1}{2(4\pi)^{3/2}}
 \int_Md^3x\sqrt g
 \int_{1/\Lambda^2}^{\infty}dt\,
 t^{-5/2}e^{-t\sigma(x)}
 \nonumber\\
 &=-\frac{\Lambda^3}{24\pi^{3/2}}
   \int_Md^3x\sqrt g
   +\frac{\Lambda}{8\pi^{3/2}}
   \int_Md^3x\sqrt g\,\sigma(x)
 \nonumber\\
 &\hspace{1.2cm}
   -\frac{1}{12\pi}
   \int_Md^3x\sqrt g\,\sigma(x)^{3/2}
   +O(\Lambda^{-1}).
 \label{eq:wdiv}
\end{align}
The linear divergence is removed by the critical tuning of the bare
mass-to-coupling ratio.  Indeed, the zero-temperature gap condition gives
\begin{equation}
 \left.\frac{\delta\Gamma}{\delta\sigma(x)}
 \right|_{T=0,\,\sigma=0}=0
 \quad\Longrightarrow\quad
 \frac{\mu_0^2}{\lambda_0}
 =-\frac{1}{2(4\pi)^{3/2}}
 \int_{1/\Lambda^2}^{\infty}dt\,t^{-3/2}
 =-\frac{\Lambda}{8\pi^{3/2}}.
 \label{eq:tune}
\end{equation}
The remaining cubic divergence is independent of $\sigma$ and is removed
by subtracting the vacuum functional,
\begin{equation}
 \Gamma_{\rm{det},\Lambda}^{(l=0)}[0]
 =-\frac{\Lambda^3}{24\pi^{3/2}}
 \int_Md^3x\sqrt g.
 \label{eq:vacsub}
\end{equation}
Combining \eqref{eq:wdiv}--\eqref{eq:vacsub}, the renormalized
zero-winding contribution is
\begin{align}
 \Gamma_{\rm{ren}}^{(l=0)}
 &\equiv
 \Gamma_{\rm{det},\Lambda}^{(l=0)}[\sigma]
 +\frac{\mu_0^2}{\lambda_0}
  \int_Md^3x\sqrt g\,\sigma(x)
 -\Gamma_{\rm{det},\Lambda}^{(l=0)}[0]
 \nonumber\\
 &=-\frac{1}{12\pi}
 \int_Md^3x\sqrt g\,\sigma(x)^{3/2}.
 \label{eq:wren}
\end{align}
Using $\chi=v^2\sigma$ and
$d\mu=d^3x\sqrt g/(v^2)^{3/2}$, this becomes
\begin{equation}
 F_{\rm{vac}}(\chi)=-\frac{\chi^{3/2}}{12\pi}.
 \label{eq:wvac}
\end{equation}
The nonzero-winding sectors are finite and give
\begin{align}
 F_{\rm{th}}(\chi)
 &=-\frac{1}{2\pi}\sum_{l=1}^{\infty}
 e^{-\ell \sqrt{\chi}}\left(\frac{\sqrt{\chi}}{l^2}+\frac{1}{l^3}\right)
 \nonumber\\
 &=-\frac{1}{2\pi}\left[
 \sqrt{\chi}\operatorname{Li}_2(e^{-\sqrt{\chi}})
 +\operatorname{Li}_3(e^{-\sqrt{\chi}})\right].
 \label{eq:wth}
\end{align}
Thus $F=F_{\rm{vac}}+F_{\rm{th}}$ is finite before the gap equation
is imposed.  The calculation above contains the only divergence involving
the dynamical field $\sigma$.  At two derivatives there can additionally be
a power-divergent term depending only on the external geometry; it is removed by a local background gravitational counterterm and does not enter the saddle equation.  The four-derivative terms are ultraviolet finite in three dimensions, so no new matter counterterm is required at the order of
interest.

Substituting \eqref{eq:hkall} into \eqref{eq:sch}, the proper-time
integrals at four derivatives are elementary.  It is convenient to define
\begin{equation}
 A(\chi)=\sum_{n\in\mathbb Z}\frac{1}{k_n^2+\chi},
 \qquad
 B(\chi)=\sum_{n\in\mathbb Z}
          \frac{k_n^2}{(k_n^2+\chi)^2},
 \qquad
 D(\chi)=\sum_{n\in\mathbb Z}
          \frac{k_n^4}{(k_n^2+\chi)^3}.
 \label{eq:sums}
\end{equation}
These thermal sums are exact and may be expressed as
\begin{equation}
 A(\chi)=\frac{1}{2\sqrt{\chi}}\coth\frac{\sqrt{\chi}}{2},\;\;
 B(\chi)=A+\chi A',\;\;
 D(\chi)=A+2\chi A'+\frac{\chi^2}{2}A'',
 \label{eq:sumid}
\end{equation}
where a prime denotes differentiation with respect to $\chi$.  Finite
temperature therefore enters through the exact discrete sums in
\eqref{eq:sums}; no zero-temperature heat-kernel approximation has been
made.

We match the result to the local hydrostatic expansion
\begin{align}
 \Gamma[\sigma]
=\int d\mu\,\Big[
 &F(\chi)+F_s(\chi)s+F_X(\chi)X
 +F_{ss}(\chi)s^2+F_{sX}(\chi)sX
 \nonumber\\
 &+F_{XX}(\chi)X^2+F_Y(\chi)Y
 +O(\partial^6)\Big].
 \label{eq:hydro}
\end{align}
using the geometric dictionary
\begin{equation}
 b^2=4s, \;\; R_{\hat h}=4X+2s,\;\;
 (\nabla_{\hat h}b)^2=4Y.
 \label{eq:dict2}
\end{equation}
Combining \eqref{eq:wvac} and \eqref{eq:wth}, and then using the
two-derivative terms in the first line of \eqref{eq:hkall}, gives
\begin{equation}
\begin{split}
 F(\chi)&=-\frac{1}{2\pi}\left[
 \operatorname{Li}_3\left(e^{-\sqrt{\chi}}\right)
 +\sqrt{\chi}\operatorname{Li}_2\left(e^{-\sqrt{\chi}}\right)
 +\frac{\chi^{3/2}}{6}
 \right],\\
 F_X(\chi)&=-\frac{1}{6}F'(\chi),
 \;\;F'(\chi)=-\frac{1}{4\pi}\log\left(2\sinh\frac{\sqrt{\chi}}{2}\right),\\
 F_s(\chi)&=-\frac{1}{3}F'(\chi)
 -\frac{\sqrt{\chi}}{24\pi}\coth\frac{\sqrt{\chi}}{2}.
\end{split}
\label{eq:wlow}
\end{equation}
they follow from the same thermal heat kernel
\eqref{eq:hkall}.

For the four-derivative terms, the proper-time integrals give the local
density
\begin{equation}
\begin{split}
 \mathcal I_4=-\frac{1}{8\pi}\Bigg\{&
 A\left(\frac{7R_{\hat{h}}^2}{1920}
       +\frac{R_{\hat{h}}b^2}{384}+\frac{b^4}{512}\right)\\
 &+B\left(-\frac{7R_{\hat{h}}b^2}{240}
       -\frac{b^4}{96}+\frac{(\nabla_{\hat{h}}b)^2}{60}\right)
 +\frac{7D}{180}b^4\Bigg\}.
\end{split}
\label{eq:i4}
\end{equation}
Equations \eqref{eq:dict} and \eqref{eq:i4} then give the off-shell
coefficient functions
\begin{equation}
\begin{split}
 F_{ss}(\chi)
 &=-\frac{1}{8\pi}
   \left(\frac{A}{15}-\frac{2B}{5}+\frac{28D}{45}\right),\;\; F_Y(\chi)=-\frac{B}{120\pi}\\
 F_{sX}(\chi)
 &=-\frac{A}{80\pi}+\frac{7B}{120\pi},
 \;\;\;F_{XX}(\chi)=-\frac{7A}{960\pi}.
\end{split}
\label{eq:woff}
\end{equation}
We next put the result on the large-$N$ saddle.  The leading gap equation
$F'(\chi_0)=0$ gives
\begin{equation}
 \chi_0=c^2,\qquad c=2\log\phi_g,
 \qquad\phi_g=\frac{1+\sqrt5}{2}.
 \label{eq:gap}
\end{equation}
At this point
\begin{align}
 A_0&=\frac{\sqrt5}{2c},
 &
 B_0&=\frac{\sqrt5}{4c}-\frac12,
 &
 D_0&=\frac{3\sqrt5}{16c}-\frac58+\frac{\sqrt5c}{8},
 \nonumber\\
 F''_0&=-\frac{\sqrt5}{16\pi c},
 &
 F'_{X,0}&=\frac{\sqrt5}{96\pi c},
 &
 F'_{s,0}&=\frac{1}{24\pi}.
 \label{eq:sadval}
\end{align}
The two-derivative correction to the saddle follows from the gap equation,
\begin{equation}
 \chi_2=-\frac{F'_{s,0}s+F'_{X,0}X}{F''_0}=\frac{X}{6}+\frac{2c}{3\sqrt5}\,s.
 \label{eq:chi2}
\end{equation}
No explicit solution for $\chi_4$ is required: because $F'(\chi_0)=0$,
it cannot contribute to the on-shell action through four derivatives.

Eliminating $\chi_2$ shifts the four-derivative coefficients according to
\begin{align}
 C_{ss}&=F_{ss,0}-\frac{(F'_{s,0})^2}{2F''_0},
 &
 C_{sX}&=F_{sX,0}-\frac{F'_{s,0}F'_{X,0}}{F''_0},
 \nonumber\\
 C_{XX}&=F_{XX,0}-\frac{(F'_{X,0})^2}{2F''_0},
 &
 C_Y&=F_{Y,0}.
 \label{eq:cdef}
\end{align}
Their explicit values are
\begin{align}
 C_{ss}
 &=\frac{1}{\pi}\left(
   \frac{17}{720}-\frac{\sqrt5}{160c}
   -\frac{\sqrt5c}{144}\right),
 &
 C_{sX}
 &=\frac{1}{\pi}\left(
   \frac{\sqrt5}{120c}-\frac{1}{45}\right),
 \nonumber\\
 C_{XX}&=-\frac{\sqrt5}{360\pi c},
 &
 C_Y&=\frac{1}{\pi}\left(
   \frac{1}{240}-\frac{\sqrt5}{480c}\right).
 \label{eq:coef4}
\end{align}
For completeness, the lower-order coefficients on the same saddle are
\begin{equation}
 F_0=-\frac{2\zeta(3)}{5\pi},
 \qquad
 F_{s,0}=-\frac{\sqrt5c}{24\pi},
 \qquad
 F_{X,0}=0.
 \label{eq:coef02}
\end{equation}
Thus the complete on-shell local action through four derivatives is
\begin{align}
 \Gamma_{{os}}
 =\int d\mu\,\Bigg[&
 -\frac{2\zeta(3)}{5\pi}
 -\frac{\sqrt5c}{24\pi}s
 +C_{ss}s^2+C_{sX}sX+C_{XX}X^2+C_Y Y
 \Bigg]+O(\partial^6).
 \label{eq:osall}
\end{align}

The quantities in \eqref{eq:coef4} are local Wilson coefficients. The first few coefficients \eqref{eq:coef02} matches with the \cite{Mauro:2026zus}. Once
they have been fixed by the heat-kernel calculation above, they do not
have to be recomputed on another stationary background (e.g pp-wave).

In particular,
for the PP-wave metric \eqref{eq:ppmet} and \eqref{eq:ppi} and \eqref{eq:ppv}, the invaraints are
For the PP-wave background, the required invariants are
\begin{equation}\label{eq:ppiv}
 s_{\rm pp}=-\beta^{2}\frac{y^{2}}{h},\;\;
 X_{\rm pp}=\frac{\beta^{2}}{2h}(2-3y^{2}),\;\;
 Y_{\rm pp}=-\frac{\beta^{4}}{h^{2}},
 \;\;\; h=1+y^{2}.
\end{equation}
The saddle-gradient invariant is
\begin{equation}
 Q_{\chi}=\beta^{2}h(\partial_y\chi)^{2},
\end{equation}
since $\chi_0=c^2$ is constant, $Q_\chi$ starts only at
$O(\partial^6)$ and does not contribute at the order considered here.
Substitution into the four-derivative part gives
the same leading gap equation 
\begin{equation}
 \chi_{\rm{pp},0}=c^2,
 \qquad
 \chi_{\rm{pp},2}
 =\frac{X_{\rm{pp}}}{6}
 +\frac{2c}{3\sqrt5}s_{\rm{pp}}.
 \label{eq:ppsad}
\end{equation}
For $h(y)=1+y^2$, this becomes
$\beta^2 m_0(y)=c^2/(1+y^2)$.  The PP-wave action is then obtained by
evaluating the same coefficient functional on the PP-wave invariant is
\begin{align}
 \Gamma_{\rm pp,os}
 =\int d\mu_{\rm pp}\Bigg[
 &-\frac{2\zeta(3)}{5\pi}
 +\frac{\sqrt5c}{24\pi}
   \frac{\beta^2y^2}{h}
 \nonumber\\
 &+\frac{\beta^4}{\pi h^2}
 \left\{
 -\frac{1}{240}
 -\frac{\sqrt5}{1440c}
 +\frac{y^2}{45}
 -\left(\frac{7}{720}
       +\frac{\sqrt5c}{144}\right)y^4
 \right\}
 \Bigg]
 +O(\partial^6).
 \label{eq:ppos}
\end{align}
with precisely the coefficients in \eqref{eq:coef4}.

\section{Heat kernel on the PP-wave}\label{app:pphk}

Starting from the PP-wave metric \eqref{eq:ppmet}, we set $v=ix$ and
$u=\tau$, so that
\begin{equation}\label{eq:ppm}
    ds^2_{\rm pp}=-2\,du\,dv+dy^2+(1+y^2)du^2.
\end{equation}
Introducing the coordinate redefinition $\tilde v=v-u/2$ gives
\begin{equation}\label{eq:brink}
     ds^2_{\rm pp}=-2\,du\,d\tilde v+dy^2+y^2du^2,
\end{equation}
which is the standard Brinkmann form \cite{Blau:2002mw} of a homogeneous plane wave.
For the thermal quotient, we identify $\tau\sim\tau+\beta$ while keeping
$x$ fixed. Since $\tilde v=ix-\tau/2$, the $n$th image in Brinkmann
coordinates is
\begin{equation}
(u,\tilde v,y)\mapsto
\left(u+n\beta,\tilde v-\frac{n\beta}{2},y\right).
\end{equation}
Thus, although the original $x$-coordinate does not shift, the transformation
to Brinkmann coordinates induces a shift in $\tilde v$.

$\Sigma(X,X')$ is Synge's world function, obtained by
evaluating the geodesic action
\begin{equation}
\Sigma(X,X')=\frac12\int_0^1 d\lambda\,
\left(-2\dot u\dot{\tilde v}+\dot y^2+y^2\dot u^2\right).
\end{equation}
Writing $a=u'-u$, the geodesic equations give
\begin{equation}
u(\lambda)=u+a\lambda,\qquad
y_{\rm cl}(\lambda)=
\frac{y\sinh[a(1-\lambda)]+y'\sinh(a\lambda)}{\sinh a}.
\end{equation}
Using $\ddot y_{\rm cl}=a^2y_{\rm cl}$, the transverse action reduces
to a boundary term,
\begin{equation}
\frac12\int_0^1d\lambda\,
\left(\dot y_{\rm cl}^{\,2}+a^2y_{\rm cl}^{\,2}\right)
=\frac12\left[y_{\rm cl}\dot y_{\rm cl}\right]_0^1.
\end{equation}
Consequently,
\begin{equation}
\Sigma(X,X')
=-a(\tilde v'-\tilde v)
+\frac{a}{2\sinh a}
\left[(y^2+y'^2)\cosh a-2yy'\right].
\end{equation}

The square root of the Van Vleck--Morette determinant in three dimensions is
\begin{equation}
    \Delta^{1/2}(X,X') = \sqrt{\frac{a}{\sinh a}}.
\end{equation}
Hence the scalar heat kernel of the PP-wave Laplacian is
\begin{equation}
    K_{\widetilde{M}}(X,X';s) = \frac{1}{(4\pi s)^{3/2}} \sqrt{\frac{a}{\sinh a}} \exp\left[ -\frac{\Sigma(X,X')}{2s} \right].
\end{equation}
We now evaluate this expression on the thermal images. For the $n$-th image, $a = n\beta$, $y'=y$  and $\tilde{v}'-\tilde{v} = -\frac{n\beta}{2}$. Therefore, 
\begin{equation}
\begin{split}
    \Sigma_n(y) &= -(n\beta)\left(-\frac{n\beta}{2}\right) + \frac{n\beta}{2\sinh(n\beta)} \left[ 2y^2\cosh(n\beta) - 2y^2 \right] \\
&= \frac{n^2\beta^2}{2} + n\beta \, y^2 \frac{\cosh(n\beta)-1}{\sinh(n\beta)} = \frac{n^2\beta^2}{2} + n\beta \, y^2 \tanh\left(\frac{n\beta}{2}\right).
\end{split}
\end{equation}

Thus the $n$-th image contribution to the diagonal thermal kernel is
\begin{equation}\label{eq:nker}
    K_n(y;s) = \frac{1}{(4\pi s)^{3/2}} \sqrt{\frac{n\beta}{\sinh(n\beta)}} \exp\left[ -\frac{n^2\beta^2}{4s} - \frac{n\beta \, y^2}{2s} \tanh\left(\frac{n\beta}{2}\right) \right].
\end{equation}

\subsection{Worldline expansion}
For the $n$th thermal image, set
$a=n\beta$, $s=\beta^2\rho$, and $\widetilde m(y)=\beta^2m(y)$.
The potential-dependent part of the worldline kernel is
\begin{equation}\label{eq:tn}
    \mathcal{T}_n[m]\equiv\left\langle
    \exp\left[-s\int_0^1dt\,m\bigl(y(t)\bigr)\right]\right\rangle=
    \left\langle\exp\left[-\rho\int_0^1dt\,
    \widetilde m\bigl(y(t)\bigr)\right]\right\rangle .
\end{equation}
Here $\langle\cdots\rangle$ denotes the normalized Gaussian worldline
average. We decompose the path into the free geodesic bridge and a Dirichlet
fluctuation. The bridge is not the saddle of the full potential-dependent
worldline action:
\begin{equation}
    y(t)=\bar y(t)+\eta(t),\;\; \eta(0)=\eta(1)=0,\implies\bar y(t)=y\,\frac{\cosh\left[a\left(t-\frac12\right)\right]}{\cosh\left(\frac a2\right)}.
    \label{eq:path}
\end{equation}
The fluctuation measure is
\begin{equation}
    \left\langle\mathcal O[\eta]\right\rangle_a=
    \frac{\displaystyle\int_{\eta(0)=\eta(1)=0}\mathcal D\eta\,\mathcal O[\eta]\,
    \exp\left[ -\frac{1}{4\beta^2\rho} \int_0^1dt\, \left(\dot\eta^2+a^2\eta^2\right)\right]}{
    \displaystyle
    \int_{\eta(0)=\eta(1)=0}\mathcal D\eta\,\exp\left[ -\frac{1}{4\beta^2\rho}
        \int_0^1dt\,\left(\dot\eta^2+a^2\eta^2\right) \right] }.
    \label{eq:gav}
\end{equation}

The Gaussian fluctuation has zero mean $  \langle\eta(t)\rangle_a=0$, and covariance
\begin{equation}
    \begin{split}
    C_a(t,t')&\equiv\left\langle\eta(t)\eta(t')\right\rangle_a
    \\
    &= 2\beta^2\rho\,\frac{\sinh(a t_<)\,\sinh\left[a(1-t_>)\right]}{a\sinh a},
    \end{split}
    \label{eq:cov}
\end{equation}
where
\begin{equation}
    t_<\equiv\min(t,t'),
    \qquad
    t_>\equiv\max(t,t').
\end{equation}
The expression is regular at $a=0$, with
\begin{equation}\label{eq:c0}
    C_0(t,t')=2\beta^2\rho\,t_<\left(1-t_>\right).
\end{equation}

The classical bridge admits an even expansion in $a=n\beta$:
\begin{equation}
    \bar y(t)=y\sum_{k=0}^{\infty}Y_k(t)(n\beta)^{2k},\;\; Y_k(t)=\frac{1}{(2k)!}\sum_{i=0}^{k}\binom{2k}{2i}\frac{E_{2i}}{2^{2i}}\left(t-\frac12\right)^{2k-2i}.
    \label{eq:bre}
\end{equation}
where $E_{2i}$ are the Euler numbers. The first few coefficients are
\begin{equation}
    Y_0(t)=1,\qquad Y_1(t)=-\frac12t(1-t),\qquad Y_2(t)=\frac{1}{24}t(1-t)\left(1+t-t^2\right).
\end{equation}

It is useful to define the bridge displacement
\begin{equation}
    \Delta\bar y(t)\equiv\bar y(t)-y
    =\sum_{k=1}^{\infty}\beta^{2k}b_k(t),\qquad
    b_k(t)\equiv n^{2k}yY_k(t).
    \label{eq:bdef}
\end{equation}

Similarly, define the covariance coefficients
$\mathscr G_q(t,t')$ by
\begin{equation}
    \frac{\sinh(a t_<)\,\sinh\left[a(1-t_>)\right]}{a\sinh a}=\sum_{q=0}^{\infty}a^{2q}\mathscr G_q(t,t').
    \label{eq:gdef}
\end{equation}
Then
\begin{equation}
     C_a(t,t')=2\rho\beta^2\sum_{q=0}^{\infty}n^{2q}\beta^{2q}\mathscr G_q(t,t').
    \label{eq:cexp}
\end{equation}
The leading coefficient is the ordinary Dirichlet Green function,
\begin{equation}
    \mathscr G_0(t,t') =t_<\left(1-t_>\right).
\end{equation}
The first correction is
\begin{equation}
    \mathscr G_1(t,t')=\frac{t_<\left(1-t_>\right)}{6}
    \left[t_<^2+\left(1-t_>\right)^2-1 \right].
\end{equation}
Equivalently, the higher coefficients may be generated recursively from
\begin{equation}
    \mathscr G_q(t,t')= (-1)^q \int_0^1du_1\cdots\int_0^1du_q\, \mathscr G_0(t,u_1) \mathscr G_0(u_1,u_2) \cdots \mathscr G_0(u_q,t').
    \label{eq:grec}
\end{equation}

Taylor expanding the potential around the classical bridge gives
\begin{equation}
    \widetilde m\bigl(\bar y(t)+\eta(t)\bigr)=
    \sum_{r=0}^{\infty}\frac{\widetilde m^{(r)}\bigl(\bar y(t)\bigr) }{r!}\,\eta(t)^r.
    \label{eq:tay}
\end{equation}
We define $    D_r(t) \equiv\widetilde m^{(r)}\bigl(\bar y(t)\bigr)$.
Since $D_0(t)$ is independent of the Gaussian fluctuation, it can be
factored out of the expectation value:
\begin{equation}
    \begin{split}
    \mathcal T_n[m]={}&\exp\left[-\rho\int_0^1dt\,D_0(t)\right]
    \left\langle\exp\left[-\rho\sum_{r=1}^{\infty}\frac{1}{r!}\int_0^1dt\,D_r(t)\eta(t)^r\right]\right\rangle_a
    \end{split}
    \label{eq:tfac}
\end{equation}
Expanding the second exponential in its number $V$ of vertices yields
\begin{equation}
     \begin{aligned}
    \mathcal T_n[m]
    ={}&
    e^{-\rho\int_0^1dt\,D_0(t)}
    \sum_{V=0}^{\infty}
    \frac{(-\rho)^V}{V!}\prod_{i=1}^{V}\int_0^1dt_i\sum_{k_1=1}^{\infty}\cdots\sum_{k_V=1}^{\infty}\left[\prod_{i=1}^{V}\frac{D_{k_i}(t_i)}{k_i!}
    \right]
    \left\langle
        \prod_{i=1}^{V}
        \eta(t_i)^{k_i}
    \right\rangle_a .
    \end{aligned}
    \label{eq:vexp}
\end{equation}
The $V=0$ contribution is understood to be equal to one.

Let $r_{ij}\in\mathbb N_0$ denote the number of Wick contractions
between vertices $i$ and $j$. For $i=j$, the integer $r_{ii}$ counts
self-contractions at vertex $i$. The allowed contractions
satisfy
\begin{equation}
      2r_{ii}+\sum_{\substack{j=1\\j\neq i}}^{V} r_{\min(i,j),\max(i,j)} = k_i,
    \qquad i=1,\ldots,V.
    \label{eq:deg}
\end{equation}
For fixed integers $k_1,\ldots,k_V$, Wick's theorem gives
\begin{equation}
    \begin{aligned}
    \left\langle\prod_{i=1}^{V}\eta(t_i)^{k_i}\right\rangle_a
    ={}&\sum_{\{r_{ij}\}}^{\prime}
    \frac{\displaystyle\prod_{i=1}^{V}k_i!}{\displaystyle\prod_{i=1}^{V}2^{r_{ii}}r_{ii}!\prod_{1\leq i<j\leq V}r_{ij}!}\prod_{i=1}^{V}C_a(t_i,t_i)^{r_{ii}}\prod_{1\leq i<j\leq V}C_a(t_i,t_j)^{r_{ij}} .
    \end{aligned}
    \label{eq:wick}
\end{equation}
The prime indicates that the degree constraints
\eqref{eq:deg} are imposed. In particular, if
$\sum_i k_i$ is odd, the constrained sum is empty and the Gaussian
expectation value vanishes.

Define the degree of vertex $i$ by
\begin{equation}
    d_i\equiv2r_{ii}+\sum_{\substack{j=1\\j\neq i}}^{V}r_{\min(i,j),\max(i,j)}.
    \label{eq:vdeg}
\end{equation}
The factors $d_i!$ in Wick's theorem cancel the Taylor factors
$1/k_i!$ in \eqref{eq:vexp}. Therefore the sums over the
$k_i$ may be eliminated completely. The resulting exact multigraph
representation is
\begin{equation}
    \begin{aligned}
    \mathcal T_n[m]
    ={}&e^{-\rho\int_0^1dt\,\widetilde m(\bar y(t))}\sum_{V=0}^{\infty}\frac{(-\rho)^V}{V!}
    \prod_{i=1}^{V}\int_0^1dt_i\sum_{\substack{r_{ij}\geq0\\d_i\geq1}}
    \left[\prod_{i=1}^{V}\widetilde m^{(d_i)}\bigl(\bar y(t_i)\bigr)\right]
    \\
    &\qquad\qquad\times\quad
    \frac{\displaystyle\prod_{i=1}^{V}C_a(t_i,t_i)^{r_{ii}}\prod_{1\leq i<j\leq V}C_a(t_i,t_j)^{r_{ij}}}{\displaystyle\prod_{i=1}^{V}2^{r_{ii}}r_{ii}!\prod_{1\leq i<j\leq V}r_{ij}!}.
    \end{aligned}
    \label{eq:multi}
\end{equation}
Thus every term is represented by a labeled Gaussian graph:

\begin{itemize}
    \item a vertex $i$ contributes
    $\widetilde m^{(d_i)}(\bar y(t_i))$;
    \item a self-edge at $i$ contributes $C_a(t_i,t_i)$;
    \item an edge between $i$ and $j$ contributes $C_a(t_i,t_j)$;
    \item multiple edges and self-edges are allowed;
    \item the condition $d_i\geq1$ excludes isolated vertices.
\end{itemize}

The factor $1/V!$, together with the edge factorials, supplies the
correct symmetry factors. Equation
\eqref{eq:multi} includes both connected and
disconnected multigraphs.

By the linked-cluster theorem, the logarithm of the potential average
is obtained by retaining only connected multigraphs:
\begin{equation}
    \begin{aligned}
    &\log\mathcal T_n[m]
    =
    -\rho\int_0^1dt\,
    \widetilde m(\bar y(t))\\& +\sum_{V=1}^{\infty}\frac{(-\rho)^V}{V!} \prod_{i=1}^{V}\int_0^1dt_i\sum_{\substack{r_{ij}\geq0,\ d_i\geq1\\\text{graph connected}}}
    \left[\prod_{i=1}^{V}\widetilde m^{(d_i)}\bigl(\bar y(t_i)\bigr)\right]
    \frac{\displaystyle\prod_{i=1}^{V}C_a(t_i,t_i)^{r_{ii}}\prod_{1\leq i<j\leq V}C_a(t_i,t_j)^{r_{ij}}}{\displaystyle\prod_{i=1}^{V}2^{r_{ii}}r_{ii}!\prod_{1\leq i<j\leq V}r_{ij}!}.
    \end{aligned}
    \label{eq:conn}
\end{equation}
This form is particularly convenient for the generation of symbolic graphs.
The disconnected terms in $\mathcal T_n$ are recovered automatically
by exponentiating the connected result.

The bridge dependence of each vertex may itself be expanded about the
common endpoint $y$. Using \eqref{eq:bdef},
\begin{equation}
    \widetilde m^{(d_i)}
    \bigl(\bar y(t_i)\bigr)=\sum_{H_i=0}^{\infty}
    \frac{\widetilde m^{(d_i+H_i)}(y)}{H_i!}
    \left[\Delta\bar y(t_i)\right]^{H_i}.
\end{equation}
Introducing occupation numbers $h_{ik}\in\mathbb N_0$ gives the more
useful form
\begin{equation}
   \begin{aligned}
    \widetilde m^{(d_i)}
    \bigl(\bar y(t_i)\bigr)
    ={}&\sum_{\{h_{ik}\}_{k\geq1}}\widetilde m^{(d_i+H_i)}(y)\prod_{k=1}^{\infty}
    \frac{\left[\beta^{2k}b_k(t_i)\right]^{h_{ik}}}{h_{ik}!},
    \end{aligned}
    \label{eq:vbri}
\end{equation}
where
\begin{equation}
    H_i=\sum_{k=1}^{\infty}h_{ik}.
\end{equation}
The bridge expansion carried by all potential vertices has total
weighted order
\begin{equation}
 S\equiv\sum_{i=1}^{V}\sum_{k=1}^{\infty}k\,h_{ik}.
    \label{eq:sord}
\end{equation}
Consequently, the vertex displacements contribute $\beta^{2S}$.

The zero-fluctuation part is expanded as
\begin{equation}
    \int_0^1dt\,\widetilde m(\bar y(t))=\widetilde m(y)+\sum_{R=1}^{\infty}\beta^{2R}M_R(y;n),
    \label{eq:mexp}
\end{equation}
where
\begin{equation}
   \begin{aligned}
    M_R(y;n)={}&\sum_{\substack{u_1,u_2,\ldots\geq0\\\sum_{k\geq1}ku_k=R}}
    \frac{\widetilde m^{(U)}(y)}{\displaystyle\prod_{k\geq1}u_k!}
    \int_0^1dt\,\prod_{k\geq1}b_k(t)^{u_k},\qquad U\equiv\sum_{k\geq1}u_k .
    \end{aligned}
    \label{eq:mr}
\end{equation}
For example,
\begin{equation}
    M_1(y;n)=-\frac{n^2y}{12}\,\widetilde m'(y),\quad  M_2(y;n)= n^4
    \left[\frac{y}{120}\widetilde m'(y)+\frac{y^2}{240}\widetilde m''(y)\right].
\end{equation}
The exponential of these corrections is
\begin{equation}
    \begin{aligned}
    &\exp\left[-\rho\int_0^1dt\,\widetilde m(\bar y(t))\right]
    =e^{-\rho\widetilde m(y)}
    \sum_{z_1,z_2,\ldots\geq0}\prod_{R=1}^{\infty}\frac{
        \left[-\rho\beta^{2R}M_R(y;n)\right]^{z_R}}{z_R!}.
    \end{aligned}
    \label{eq:zexp}
\end{equation}
with weighted order
\begin{equation}
   p\equiv\sum_{R=1}^{\infty}Rz_R.
    \label{eq:pord}
\end{equation}
A term with occupation numbers $\{z_R\}$ therefore contributes
$\beta^{2p}$.

For every pair $i\leq j$, let $N_{ij}^{(q)}$ count the number of
propagators for which the coefficient $\mathscr G_q$ in
\eqref{eq:cexp} has been selected. Thus, the number of Wick contractions is
\begin{equation}
    r_{ij}=\sum_{q=0}^{\infty}N_{ij}^{(q)}.
\end{equation}
Defining $\kappa_{ij}\equiv2-\delta_{ij}$, the propagator factors may
be expanded uniformly as
\begin{equation}
    \begin{aligned}
    \frac{C_a(t_i,t_j)^{r_{ij}}}{2^{\delta_{ij}r_{ij}}r_{ij}!}
    ={}&\sum_{\substack{N_{ij}^{(q)}\geq0\\\sum_qN_{ij}^{(q)}=r_{ij}}}
    \prod_{q=0}^{\infty}\frac{1}{N_{ij}^{(q)}!}
    \left[\kappa_{ij}\rho\,n^{2q}\mathscr G_q(t_i,t_j)\beta^{2(q+1)}\right]^{N_{ij}^{(q)}}.
    \end{aligned}
    \label{eq:prop}
\end{equation}
The total number of propagators is
\begin{equation}
    P\equiv\sum_{1\leq i\leq j\leq V}\sum_{q=0}^{\infty}N_{ij}^{(q)}=\sum_{1\leq i\leq j\leq V}r_{ij},
    \label{eq:pnum}
\end{equation}
while the additional covariance order is
\begin{equation}
    Q\equiv\sum_{1\leq i\leq j\leq V}\sum_{q=0}^{\infty}q\,N_{ij}^{(q)}.
    \label{eq:qord}
\end{equation}
The complete propagator product therefore contributes
\begin{equation}
    \beta^{2(P+Q)}.
\end{equation}

Combining the propagator order, the bridge corrections at the
vertices, and the expansion of the zero-fluctuation exponential, the
total worldline order is
\begin{equation}
    R_{{WL}}
    =P+Q+S+p.
    \label{eq:wlord}
\end{equation}
Therefore a contribution belongs to the coefficient of
$\beta^{2R}$ precisely when
\begin{equation}
    P+Q+S+p=R.
    \label{eq:fixord}
\end{equation}

This counting is finite at every fixed order. Indeed, the degree
constraints imply
\begin{equation}
    \sum_{i=1}^{V}d_i =2\sum_{1\leq i\leq j\leq V}r_{ij}=2P.
    \label{eq:dsum}
\end{equation}
Because every vertex has $d_i\geq1$, one has $V\leq2P$; moreover,
\eqref{eq:fixord} implies $P\leq R$, and hence $V\leq2R$. Thus only
finitely many vertices, propagators, bridge insertions, and covariance
corrections can contribute to a prescribed power $\beta^{2R}$.

The connected worldline graphs define
\begin{equation}
    \mathcal T_n[m]
    =e^{-\rho \widetilde m(y)}
    \exp\left[\beta^2\ell_2+\beta^4\ell_4+\beta^6\ell_6+O(\beta^8)\right].
\label{eq:tcon}
\end{equation}
In the formulas below, $\widetilde m^{(r)}$ denotes the $r$th derivative with respect
to $y$, evaluated at the displayed point $y$.  The first connected
coefficient is
\begin{equation}
    \ell_2=\frac{\rho n^2y}{12}\widetilde m'-\frac{\rho^2}{6}\widetilde m''+\frac{\rho^3}{12}(\widetilde m')^2.
\label{eq:l2}
\end{equation}
At fourth order one finds
\begin{equation}
\begin{aligned}
    \ell_4={}&
    -\frac{\rho n^4y}{120}\widetilde m'-\frac{\rho n^4y^2}{240}\widetilde m''+\frac{\rho^2n^2}{90}\widetilde m''+\frac{\rho^2n^2y}{60}\widetilde m^{(3)} -\frac{\rho^3}{60}\widetilde m^{(4)}
    -\frac{\rho^3n^2}{120}(\widetilde m')^2
    \\
    &
    -\frac{\rho^3n^2y}{60}\widetilde m'\widetilde m''+\frac{\rho^4}{30}\widetilde m'\widetilde m^{(3)}+\frac{\rho^4}{90}(\widetilde m'')^2
    -\frac{\rho^5}{60}(\widetilde m')^2\widetilde m''.
\end{aligned}
\label{eq:l4}
\end{equation}
The complete sixth-order connected coefficient is
\begin{equation}
\begin{split}\label{eq:l6}
    \ell_6 ={}& \rho n^6\left( \frac{17y\widetilde{m}'}{20160} +\frac{17y^2\widetilde{m}''}{20160} +\frac{y^3\widetilde{m}^{(3)}}{6720} \right) +\rho^2n^4\left( -\frac{\widetilde{m}''}{945} -\frac{29y\widetilde{m}^{(3)}}{10080} -\frac{y^2\widetilde{m}^{(4)}}{1120} \right) \\
    &+\rho^3n^2\left( \frac{\widetilde{m}^{(4)}}{420} +\frac{y\widetilde{m}^{(5)}}{560} \right) 
    +\rho^3n^4\left( \frac{17(\widetilde{m}')^2}{20160} +\frac{17y\widetilde{m}'\widetilde{m}''}{5040} +\frac{y^2\widetilde{m}'\widetilde{m}^{(3)}}{1120} +\frac{17y^2(\widetilde{m}'')^2}{20160} \right) \\
    & -\frac{\rho^4}{840}\widetilde{m}^{(6)}  +\rho^4n^2\left( -\frac{29\widetilde{m}'\widetilde{m}^{(3)}}{5040} -\frac{2(\widetilde{m}'')^2}{945} -\frac{y\widetilde{m}'\widetilde{m}^{(4)}}{280} -\frac{29y\widetilde{m}''\widetilde{m}^{(3)}}{5040} \right) \\
    & +\rho^5\left( \frac{\widetilde{m}'\widetilde{m}^{(5)}}{280} +\frac{\widetilde{m}''\widetilde{m}^{(4)}}{210} +\frac{23(\widetilde{m}^{(3)})^2}{5040} \right) +\rho^5n^2\left( \frac{17(\widetilde{m}')^2\widetilde{m}''}{5040} +\frac{y(\widetilde{m}')^2\widetilde{m}^{(3)}}{560} +\frac{17y\widetilde{m}'(\widetilde{m}'')^2}{5040} \right) \\
    & +\rho^6\left( -\frac{(\widetilde{m}')^2\widetilde{m}^{(4)}}{280} -\frac{29\widetilde{m}'\widetilde{m}''\widetilde{m}^{(3)}}{2520} -\frac{4(\widetilde{m}'')^3}{2835} \right) +\rho^7\left( \frac{(\widetilde{m}')^3\widetilde{m}^{(3)}}{840} +\frac{17(\widetilde{m}')^2(\widetilde{m}'')^2}{5040} \right).
\end{split}
\end{equation}
These expressions contain respectively $3$, $10$, and $28$ collected
connected structures. 

Writing
\begin{equation}
    \mathcal T_n[m]
    =e^{-\rho\widetilde m(y)}
    \left[
        1+\beta^2A_2+\beta^4A_4+\beta^6A_6
        +O(\beta^8)
    \right],
\end{equation}
the linked-cluster theorem gives
\begin{equation}
    A_2=\ell_2,\qquad   A_4=\ell_4+\frac12\ell_2^2,\qquad  A_6
    =\ell_6+\ell_2\ell_4+\frac16\ell_2^3.
\end{equation}
Multiplying the potential average by the geometrical kernel gives the full
coefficients appearing in \eqref{eq:hkppow}:
\begin{equation}\label{eq:bcoef}
    B_0=1,\qquad  B_2=q_2+A_2,\qquad  B_4=q_4+A_4+q_2A_2,\qquad  B_6=q_6+A_6+q_2A_4+q_4A_2.
\end{equation}

\section{Low-Temperature details}\label{ap:ld}
Motivated from \eqref{eq:lsd} we have
\begin{equation}
\begin{split}
    f_\beta[m]={}&f_{{th},\beta}[0]
    +\frac12\langle m,H_0m\rangle
    +\langle J_1+J_2,m\rangle \\
    &+f_{{vac}}^{(3)}[m]
    +f_{{th}}^{(2)}[m]
    +O\left(\frac{q^4}{\beta}\right),
    \label{eq:func}
\end{split}
\end{equation}
and using $m=m_1+m_2$, we have \eqref{eq:m1m2} and every term linear in $m_2$ is proportional to
\begin{equation}
    \langle H_0m_1+J_1,m_2\rangle=0.
    \label{eq:m2canc}
\end{equation}
Thus the complete on-shell answer through $q^3$ depends only on the known
profile $S_1$.  The contributions at this order are the free gas, the
quadratic thermal response of the first winding, the coupling to the second
winding, and the cubic vacuum vertex.

To evaluate the first-winding response, set $k=\kappa/\beta$.  The scaled
transverse oscillator is
\begin{equation}
    h_\kappa=-\frac{1}{2\kappa}\frac{ d^2}{ d\eta^2}
    +\frac{\kappa\eta^2}{2},
    \qquad
    h_\kappa|a,\kappa\rangle=\left(a+\frac12\right)|a,\kappa\rangle.
    \label{eq:osc}
\end{equation}
Ordinary Rayleigh--Schr\"odinger perturbation theory gives
\begin{equation}
    E_{0,\kappa/\beta}[m_1]
    =\frac12+\frac{\kappa}{2\beta}
    +qv_\kappa+q^2w_\kappa+O(q^3),
    \label{eq:energyexp}
\end{equation}
with
\begin{equation}
\begin{split}
    v_\kappa
    &=\left\langle0,\kappa\left|
    \frac{S_1}{2\kappa}\right|0,\kappa\right\rangle,
    \\
    w_\kappa
    &=-\sum_{a=1}^{\infty}\frac{1}{a}
    \left|
    \left\langle a,\kappa\left|
    \frac{S_1}{2\kappa}\right|0,\kappa\right\rangle
    \right|^2.
    \label{eq:vwdef}
\end{split}
\end{equation}
The first-order shift can be evaluated analytically:
\begin{equation}
    v_\kappa=\frac{4}{\pi^2}
    e^{\kappa/2}E_1\left(\frac{\kappa}{2}\right),
    \qquad
    E_1(x)=\int_x^\infty\frac{e^{-t}}{t}\, dt.
    \label{eq:vanalytic}
\end{equation}
Two moments needed below are
\begin{equation}
\begin{split}
    D_1&=\int_0^\infty\frac{ d\kappa}{2\pi}
    e^{-\kappa/2}v_\kappa=\frac{4}{\pi^3},
    \\
    D_2&=\int_0^\infty\frac{ d\kappa}{2\pi}
    e^{-\kappa/2}v_\kappa^2=\frac{8}{3\pi^3}.
    \label{eq:d1d2}
\end{split}
\end{equation}
Here we used
\begin{equation}
    \int_0^\infty E_1(x)\, dx=1,
    \qquad
    \int_0^\infty e^xE_1(x)^2\, dx=\frac{\pi^2}{6}.
\end{equation}
Indeed, the change in the ground-state part of the first winding is
\begin{equation}
\begin{split}
    \Delta W_{1,0}
    &=\frac{q}{\beta}
    \int_0^\infty\frac{ d\kappa}{2\pi}e^{-\kappa/2}
    \left[e^{-\beta qv_\kappa-\beta q^2w_\kappa+\cdots}-1\right]
    \\
    &=-D_1q^2
    +\left(\frac{\beta}{2}D_2-\mathcal R_{\rm{off}}\right)q^3
    +O(q^4),
    \label{eq:w1exp}
\end{split}
\end{equation}
where
\begin{equation}
    \mathcal R_{\rm{off}}
    =\int_0^\infty\frac{ d\kappa}{2\pi}
    e^{-\kappa/2}w_\kappa.
    \label{eq:roffdef}
\end{equation}
Half of the order-$q^2$ term in
Eq.~\eqref{eq:w1exp} is cancelled by the vacuum
quadratic term, leaving $-D_1q^2/2=-2q^2/\pi^3$, as found above.  At order
$q^3$, the diagonal part gives
\begin{equation}
    \frac{\beta}{2}D_2q^3
    =\frac{4\beta}{3\pi^3}q^3.
    \label{eq:diagq3}
\end{equation}

The coupling of $m_1$ to the second thermal winding is determined by
\begin{equation}
\begin{split}
    L_2
    &=\int_{-\infty}^{\infty} d\eta\,
    j_2(\eta)S_1(\eta) \\
    &=\frac{4}{\pi^3}\int_0^\infty ds\,s
    K_0(s)K_0\left(\frac{s}{\sqrt2}\right)
    =\frac{4\log2}{\pi^3}.
    \label{eq:w2overlap}
\end{split}
\end{equation}
We used the identity
\begin{equation}
    \int_0^\infty ds\,sK_0(as)K_0(bs)
    =\frac{\log(a/b)}{a^2-b^2}.
\end{equation}
Since $W=-\beta f_{{on}}$, this term contributes $-L_2q^3$.

The cubic vacuum contribution follows from
Eq.~\eqref{eq:f3vac}.  Using
Eq.~\eqref{eq:f3vac}, all three factors of momentum cancel,
and one obtains
\begin{equation}
\begin{split}
    \beta f_{{vac}}^{(3)}[m_1]
    &=\frac{q^3}{48}\left(\frac{8}{\pi}\right)^3
    \int\frac{ ds_1 ds_2}{(2\pi)^2}
    \prod_{i=1}^{3}K_0\left(\frac{|s_i|}{\sqrt2}\right)
    \\
    &=\frac{16}{3\pi^3}q^3,
    \qquad s_3=-s_1-s_2.
    \label{eq:vactri}
\end{split}
\end{equation}
For the last equality, the convolution theorem and
\begin{equation}
    \int_{-\infty}^{\infty}\frac{ d\eta}
    {8(\eta^2+1/2)^{3/2}}=\frac12
\end{equation}
reduce the double integral to an elementary one.  The corresponding
contribution to $W$ is therefore $-16q^3/(3\pi^3)$.

\paragraph{Off-diagonal Response}

The remaining contribution is
\begin{equation}
\begin{split}
\mathcal R_{{off}}
&=
\int_0^\infty\frac{ d\kappa}{2\pi}
e^{-\kappa/2}w_\kappa,
\\
w_\kappa
&=
-\sum_{a=1}^{\infty}\frac{|M_a(\kappa)|^2}{a},
\qquad
M_a(\kappa)
=
\left\langle a,\kappa\left|
\frac{S_1}{2\kappa}\right|0,\kappa\right\rangle .
\end{split}
\label{eq:rodef}
\end{equation}
The oscillator matrix element of a plane wave is
\begin{equation}
\left\langle a,\kappa\left|
e^{is\eta}\right|0,\kappa\right\rangle
=
\frac{(is)^a}{(2\kappa)^{a/2}\sqrt{a!}}
e^{-s^2/(4\kappa)}.
\label{eq:pw}
\end{equation}
Since $S_1$ is even, only the levels $a=2r$ contribute. We use
\begin{equation}
\widetilde S_1(s)
=
\frac{8}{\pi}|s|
K_0\left(\frac{|s|}{\sqrt2}\right)
\label{eq:s1f}
\end{equation}
and the Schwinger representation
\begin{equation}
K_0\left(\frac{|s|}{\sqrt2}\right)
=
\frac12\int_0^\infty\frac{ du}{u}
\exp\left(-u-\frac{s^2}{8u}\right).
\label{eq:k0}
\end{equation}
Substituting \eqref{eq:pw}--\eqref{eq:k0} into
\eqref{eq:rodef} and performing the Gaussian $s$-integral gives
\begin{equation}
M_{2r}(2x)
=
(-1)^r
\frac{2^{r+2}r!}{\pi^2\sqrt{(2r)!}}
\int_0^\infty du \, e^{-u}
\frac{u^r}{(u+x)^{r+1}}.
\label{eq:m2r}
\end{equation}
It follows that
\begin{equation}
\mathcal R_{{off}}
=
-\frac{16}{\pi^5}
\sum_{r=1}^{\infty}
\frac{4^r(r!)^2}{(2r)!(2r)} \mathcal I_r,
\label{eq:roser}
\end{equation}
where
\begin{equation}
\mathcal I_r
=
\int_0^\infty dx \, e^{-x}
\left[
\int_0^\infty du \, e^{-u}
\frac{u^r}{(u+x)^{r+1}}
\right]^2.
\label{eq:irdef}
\end{equation}
In each inner integral, set $t=u/(u+x)$. The remaining $x$-integral is elementary, and one obtains
\begin{equation}
\begin{split}
\mathcal I_r
&=
\int_0^1 dt\int_0^1 dv \, 
\frac{(tv)^r}{1-tv}
\\
&=
\sum_{j=0}^{\infty}\frac{1}{(r+j+1)^2}
=
\zeta(2)-H_r^{(2)},
\end{split}
\label{eq:ir}
\end{equation}
with
\begin{equation}
H_r^{(2)}=\sum_{j=1}^r\frac{1}{j^2}.
\label{eq:hr}
\end{equation}
Therefore
\begin{equation}
\mathcal R_{{off}}
=
-\frac{16}{\pi^5}\mathcal S,
\qquad
\mathcal S
=
\sum_{r=1}^{\infty}
\frac{4^r(r!)^2}{(2r)!(2r)}
\left[\zeta(2)-H_r^{(2)}\right].
\label{eq:sdef}
\end{equation}
To perform this sum, use
\begin{equation}
\zeta(2)-H_r^{(2)}
=
\int_0^1 dz \, 
\frac{-z^r\log z}{1-z}.
\label{eq:hint}
\end{equation}
The required inverse-central-binomial generating function follows from
\begin{equation}
\frac{4^r(r!)^2}{(2r)!}
=
(2r+1)\int_0^1 du \, 
[4u(1-u)]^r.
\label{eq:beta}
\end{equation}
Indeed, defining $X=4zu(1-u)$, one finds
\begin{equation}
\begin{split}
F(z)
&\equiv
\sum_{r=1}^{\infty}
\frac{4^r(r!)^2}{(2r)!(2r)}z^r
\\
&=
\int_0^1 du
\left[
\frac{X}{1-X}
-\frac12\log(1-X)
\right]
\\
&=
\sqrt{\frac{z}{1-z}}\arcsin\sqrt z.
\end{split}
\label{eq:gen}
\end{equation}
Equations \eqref{eq:hint} and \eqref{eq:gen} reduce
$\mathcal S$ to
\begin{equation}
\begin{split}
\mathcal S
&=
\int_0^1 dz \, 
\frac{-\log z}{1-z}
\sqrt{\frac{z}{1-z}}\arcsin\sqrt z
\\
&=
-4\int_0^{\pi/2} d\theta \, 
\theta\tan^2\theta\log(\sin\theta),
\end{split}
\label{eq:sint}
\end{equation}
where $z=\sin^2\theta$. Integrating by parts and using
\begin{equation}
\begin{split}
\int_0^{\pi/2} d\theta \, 
\tan\theta\log(\sin\theta)
&=-\frac{\pi^2}{24},
\\
\int_0^{\pi/2} d\theta \, 
\theta\log(\sin\theta)
&=-\frac{\pi^2}{8}\log2
+\frac{7}{16}\zeta(3),
\end{split}
\label{eq:ang}
\end{equation}
gives
\begin{equation}
\mathcal S
=
\frac{\pi^2}{3}
-\frac{\pi^2}{2}\log2
+\frac74\zeta(3).
\label{eq:sval}
\end{equation}
The second identity in \eqref{eq:ang} follows directly from
$\log(2\sin\theta) = -\sum_{n=1}^{\infty}\cos(2n\theta)/n$.
Combining \eqref{eq:sdef} and \eqref{eq:sval}, we obtain the exact off-diagonal response
\begin{equation}
\mathcal R_{\rm{off}}
=
\frac{8\log2-\frac{16}{3}}{\pi^3}
-\frac{28\zeta(3)}{\pi^5}.
\label{eq:ro}
\end{equation}
Consequently, the $\beta$-independent interaction coefficient in the third exponential sector is
\begin{equation}
\begin{split}
A
&=
\frac{16}{3\pi^3}
+\frac{4\log2}{\pi^3}
+\mathcal R_{\rm{off}}
\\
&=
\frac{12\log2}{\pi^3}
-\frac{28\zeta(3)}{\pi^5}.
\end{split}
\label{eq:a}
\end{equation}
The complete $q^3$ contribution is therefore
\begin{equation}
W^{(3)}(\beta)
=
\left[
\frac{10}{9\pi\beta}
+\frac{4\beta}{3\pi^3}
-\frac{12\log2}{\pi^3}
+\frac{28\zeta(3)}{\pi^5}
\right]q^3.
\label{eq:w3}
\end{equation}

\bibliography{biblio}

\providecommand{\href}[2]{#2}\begingroup\raggedright\begin{thebibliography}{10}

\bibitem{Bhattacharyya:2007vs}
S.~Bhattacharyya, S.~Lahiri, R.~Loganayagam and S.~Minwalla, \emph{{Large rotating AdS black holes from fluid mechanics}}, \href{https://doi.org/10.1088/1126-6708/2008/09/054}{\emph{JHEP} {\bfseries 09} (2008) 054} [\href{https://arxiv.org/abs/0708.1770}{{\ttfamily 0708.1770}}].

\bibitem{Anand:2025mfh}
H.~Anand, N.~Benjamin, V.~Kumar, S.~Minwalla, J.~Mukherjee, S.~Pal et~al., \emph{{Semi-universality of CFT$_d$ entropy at large spin}}, \href{https://doi.org/10.21468/SciPostPhys.21.2.042}{\emph{SciPost Phys.} {\bfseries 21} (2026) 042} [\href{https://arxiv.org/abs/2512.00158}{{\ttfamily 2512.00158}}].

\bibitem{Bajaj:2024utv}
K.~Bajaj, V.~Kumar, S.~Minwalla, J.~Mukherjee and A.~Rahaman, \emph{{Grey Galaxies in $AdS_5$}}, \href{https://doi.org/10.21468/SciPostPhys.20.6.178}{\emph{SciPost Phys.} {\bfseries 20} (2026) 178} [\href{https://arxiv.org/abs/2412.06904}{{\ttfamily 2412.06904}}].

\bibitem{Kim:2023sig}
S.~Kim, S.~Kundu, E.~Lee, J.~Lee, S.~Minwalla and C.~Patel, \emph{{Grey Galaxies{\textquoteright} as an endpoint of the Kerr-AdS superradiant instability}}, \href{https://doi.org/10.1007/JHEP11(2023)024}{\emph{JHEP} {\bfseries 11} (2023) 024} [\href{https://arxiv.org/abs/2305.08922}{{\ttfamily 2305.08922}}].

\bibitem{Komargodski:2026ain}
Z.~Komargodski, A.~Miscioscia and F.K.~Popov, \emph{{Regge's Inferno}},  \href{https://arxiv.org/abs/2603.10197}{{\ttfamily 2603.10197}}.

\bibitem{Mukherjee:2026dfu}
J.~Mukherjee and P.~Ray, \emph{{Semi-universality of conformal higher-derivative and conformal higher-spin fields}}, \href{https://doi.org/10.1007/JHEP08(2026)182}{\emph{JHEP} {\bfseries 08} (2026) 182} [\href{https://arxiv.org/abs/2606.08461}{{\ttfamily 2606.08461}}].

\bibitem{Banerjee:2012iz}
N.~Banerjee, J.~Bhattacharya, S.~Bhattacharyya, S.~Jain, S.~Minwalla and T.~Sharma, \emph{Constraints on fluid dynamics from equilibrium partition functions}, \href{https://doi.org/10.1007/JHEP09(2012)046}{\emph{JHEP} {\bfseries 09} (2012) 046} [\href{https://arxiv.org/abs/1203.3544}{{\ttfamily 1203.3544}}].

\bibitem{Jensen:2012jh}
K.~Jensen, M.~Kaminski, P.~Kovtun, R.~Meyer, A.~Ritz and A.~Yarom, \emph{{Towards hydrodynamics without an entropy current}}, \href{https://doi.org/10.1103/PhysRevLett.109.101601}{\emph{Phys. Rev. Lett.} {\bfseries 109} (2012) 101601} [\href{https://arxiv.org/abs/1203.3556}{{\ttfamily 1203.3556}}].

\bibitem{Benjamin:2023ThermalEFT}
N.~Benjamin, J.~Lee, H.~Ooguri and D.~Simmons-Duffin, \emph{Universal asymptotics for high energy cft data}, \href{https://doi.org/10.1007/JHEP03(2024)115}{\emph{JHEP} {\bfseries 03} (2024) 115} [\href{https://arxiv.org/abs/2306.08031}{{\ttfamily 2306.08031}}].

\bibitem{Allameh:2024qqp}
K.~Allameh and E.~Shaghoulian, \emph{{Modular invariance and thermal effective field theory in CFT}}, \href{https://doi.org/10.1007/JHEP01(2025)200}{\emph{JHEP} {\bfseries 01} (2025) 200} [\href{https://arxiv.org/abs/2402.13337}{{\ttfamily 2402.13337}}].

\bibitem{PhysRevD.7.2911}
K.G.~Wilson, \emph{Quantum field - theory models in less than 4 dimensions}, \href{https://doi.org/10.1103/PhysRevD.7.2911}{\emph{Phys. Rev. D} {\bfseries 7} (1973) 2911}.

\bibitem{Moshe:2003xn}
M.~Moshe and J.~Zinn-Justin, \emph{Quantum field theory in the large {$N$} limit: A review}, \href{https://doi.org/10.1016/S0370-1573(03)00263-1}{\emph{Phys. Rept.} {\bfseries 385} (2003) 69} [\href{https://arxiv.org/abs/hep-th/0306133}{{\ttfamily hep-th/0306133}}].

\bibitem{PhysRevD.10.2491}
S.~Coleman, R.~Jackiw and H.D.~Politzer, \emph{Spontaneous symmetry breaking in the $\mathrm{O}(n)$ model for large $n$}, \href{https://doi.org/10.1103/PhysRevD.10.2491}{\emph{Phys. Rev. D} {\bfseries 10} (1974) 2491}.

\bibitem{PhysRevD.13.2212}
L.F.~Abbott, J.S.~Kang and H.J.~Schnitzer, \emph{Bound states, tachyons, and restoration of symmetry in the $\frac{1}{N}$ expansion}, \href{https://doi.org/10.1103/PhysRevD.13.2212}{\emph{Phys. Rev. D} {\bfseries 13} (1976) 2212}.

\bibitem{ChubukovSachdevYe:1994}
A.V.~Chubukov, S.~Sachdev and J.~Ye, \emph{Theory of two-dimensional quantum heisenberg antiferromagnets with a nearly critical ground state}, \href{https://doi.org/10.1103/PhysRevB.49.11919}{\emph{Phys. Rev. B} {\bfseries 49} (1994) 11919} [\href{https://arxiv.org/abs/cond-mat/9304046}{{\ttfamily cond-mat/9304046}}].

\bibitem{Sachdev:1993pr}
S.~Sachdev, \emph{{Polylogarithm identities in a conformal field theory in three-dimensions}}, \href{https://doi.org/10.1016/0370-2693(93)90935-B}{\emph{Phys. Lett. B} {\bfseries 309} (1993) 285} [\href{https://arxiv.org/abs/hep-th/9305131}{{\ttfamily hep-th/9305131}}].

\bibitem{Romatschke:2019ybu}
P.~Romatschke, \emph{{Finite-Temperature Conformal Field Theory Results for All Couplings: O(N) Model in 2+1 Dimensions}}, \href{https://doi.org/10.1103/PhysRevLett.122.231603}{\emph{Phys. Rev. Lett.} {\bfseries 122} (2019) 231603} [\href{https://arxiv.org/abs/1904.09995}{{\ttfamily 1904.09995}}].

\bibitem{Petkou:1998fc}
A.C.~Petkou and N.D.~Vlachos, \emph{{Finite size and finite temperature effects in the conformally invariant O(N) vector model for 2 less than d less than 4}},  in \emph{{5th International Workshop on Thermal Field Theories and Their Applications}}, 9, 1998 [\href{https://arxiv.org/abs/hep-th/9809096}{{\ttfamily hep-th/9809096}}].

\bibitem{Iliesiu:2018fao}
L.~Iliesiu, M.~Kolo{\u{g}}lu, R.~Mahajan, E.~Perlmutter and D.~Simmons-Duffin, \emph{{The Conformal Bootstrap at Finite Temperature}}, \href{https://doi.org/10.1007/JHEP10(2018)070}{\emph{JHEP} {\bfseries 10} (2018) 070} [\href{https://arxiv.org/abs/1802.10266}{{\ttfamily 1802.10266}}].

\bibitem{David:2024pir}
J.R.~David and S.~Kumar, \emph{{The large N vector model on S$^{1}$ {\texttimes} S$^{2}$}}, \href{https://doi.org/10.1007/JHEP03(2025)169}{\emph{JHEP} {\bfseries 03} (2025) 169} [\href{https://arxiv.org/abs/2411.18509}{{\ttfamily 2411.18509}}].

\bibitem{David:2025tqn}
J.R.~David and S.~Kumar, \emph{{High to low temperature: O(N) model at large N}}, \href{https://doi.org/10.1007/JHEP02(2026)194}{\emph{JHEP} {\bfseries 02} (2026) 194} [\href{https://arxiv.org/abs/2508.14872}{{\ttfamily 2508.14872}}].

\bibitem{Mauro:2026zus}
L.~Mauro and A.~Vichi, \emph{{Thermal effective action for the $O(N)$ vector model}},  \href{https://arxiv.org/abs/2606.05059}{{\ttfamily 2606.05059}}.

\bibitem{David:2026yis}
J.R.~David and S.~Kumar, \emph{{The large $N$ vector model with angular velocity}},  \href{https://arxiv.org/abs/2608.31151}{{\ttfamily 2608.31151}}.

\bibitem{Tolman:1930gr}
R.C.~Tolman and P.~Ehrenfest, \emph{Temperature equilibrium in a static gravitational field}, \href{https://doi.org/10.1103/PhysRev.36.1791}{\emph{Phys. Rev.} {\bfseries 36} (1930) 1791}.

\bibitem{Giombi:2019upv}
S.~Giombi, R.~Huang, I.R.~Klebanov, S.S.~Pufu and G.~Tarnopolsky, \emph{{The $O(N)$ Model in $4<d<6$: Instantons and Complex CFTs}}, \href{https://doi.org/10.1103/PhysRevD.101.045013}{\emph{Phys. Rev. D} {\bfseries 101} (2020) 045013} [\href{https://arxiv.org/abs/1910.02462}{{\ttfamily 1910.02462}}].

\bibitem{Advant:2026cqt}
A.~Advant, H.~Anand, N.~Benjamin, V.~Kumar, S.~Minwalla, J.~Mukherjee et~al., \emph{{From $Z$ to $a$: High-temperature relations, subleading semi-universality, and conformal anomalies}},  \href{https://arxiv.org/abs/2609.03013}{{\ttfamily 2609.03013}}.

\bibitem{Loganayagam:2008is}
R.~Loganayagam, \emph{{Entropy Current in Conformal Hydrodynamics}}, \href{https://doi.org/10.1088/1126-6708/2008/05/087}{\emph{JHEP} {\bfseries 05} (2008) 087} [\href{https://arxiv.org/abs/0801.3701}{{\ttfamily 0801.3701}}].

\bibitem{Schwinger:1951nm}
J.S.~Schwinger, \emph{{On gauge invariance and vacuum polarization}}, \href{https://doi.org/10.1103/PhysRev.82.664}{\emph{Phys. Rev.} {\bfseries 82} (1951) 664}.

\bibitem{Giombi:2008vd}
S.~Giombi, A.~Maloney and X.~Yin, \emph{{One-loop Partition Functions of 3D Gravity}}, \href{https://doi.org/10.1088/1126-6708/2008/08/007}{\emph{JHEP} {\bfseries 08} (2008) 007} [\href{https://arxiv.org/abs/0804.1773}{{\ttfamily 0804.1773}}].

\bibitem{Vassilevich:2003xt}
D.V.~Vassilevich, \emph{{Heat kernel expansion: User's manual}}, \href{https://doi.org/10.1016/j.physrep.2003.09.002}{\emph{Phys. Rept.} {\bfseries 388} (2003) 279} [\href{https://arxiv.org/abs/hep-th/0306138}{{\ttfamily hep-th/0306138}}].

\bibitem{Fliegner:1994zc}
D.~Fliegner, P.~Haberl, M.G.~Schmidt and C.~Schubert, \emph{{An improved heat kernel expansion from worldline path integrals}}, {\emph{Discourses Math. Appl.} {\bfseries 4} (1995) 87} [\href{https://arxiv.org/abs/hep-th/9411177}{{\ttfamily hep-th/9411177}}].

\bibitem{Diatlyk:2023msc}
O.~Diatlyk, F.K.~Popov and Y.~Wang, \emph{{Beyond~$N=\infty$ in Large $N$ conformal vector models at finite temperature}}, \href{https://doi.org/10.1007/JHEP08(2024)219}{\emph{JHEP} {\bfseries 08} (2024) 219} [\href{https://arxiv.org/abs/2309.02347}{{\ttfamily 2309.02347}}].

\bibitem{Chamati:2011zz}
H.~Chamati and N.S.~Tonchev, \emph{{Quantum critical scaling and the Gross-Neveu model in 2 + 1 dimensions}}, \href{https://doi.org/10.1209/0295-5075/95/40005}{\emph{EPL} {\bfseries 95} (2011) 40005} [\href{https://arxiv.org/abs/1112.4853}{{\ttfamily 1112.4853}}].

\bibitem{Aharony:2012ns}
O.~Aharony, S.~Giombi, G.~Gur-Ari, J.~Maldacena and R.~Yacoby, \emph{{The Thermal Free Energy in Large N Chern-Simons-Matter Theories}}, \href{https://doi.org/10.1007/JHEP03(2013)121}{\emph{JHEP} {\bfseries 03} (2013) 121} [\href{https://arxiv.org/abs/1211.4843}{{\ttfamily 1211.4843}}].

\bibitem{Jain:2013py}
S.~Jain, S.~Minwalla, T.~Sharma, T.~Takimi, S.R.~Wadia and S.~Yokoyama, \emph{{Phases of large $N$ vector Chern-Simons theories on $S^2 \times S^1$}}, \href{https://doi.org/10.1007/JHEP09(2013)009}{\emph{JHEP} {\bfseries 09} (2013) 009} [\href{https://arxiv.org/abs/1301.6169}{{\ttfamily 1301.6169}}].

\bibitem{Amsterdamski:1989bt}
P.~Amsterdamski, A.L.~Berkin and D.J.~O'Connor, \emph{{$B$(8) 'Hamidew' Coefficient for a Scalar Field}}, \href{https://doi.org/10.1088/0264-9381/6/12/024}{\emph{Class. Quant. Grav.} {\bfseries 6} (1989) 1981}.

\bibitem{Blau:2002mw}
M.~Blau, J.M.~Figueroa-O'Farrill and G.~Papadopoulos, \emph{{Penrose limits, supergravity and brane dynamics}}, \href{https://doi.org/10.1088/0264-9381/19/18/310}{\emph{Class. Quant. Grav.} {\bfseries 19} (2002) 4753} [\href{https://arxiv.org/abs/hep-th/0202111}{{\ttfamily hep-th/0202111}}].

\end{thebibliography}\endgroup
\bibliographystyle{JHEP}

\end{document}